\documentclass[useAMS,usenatbib]{mn2e}
\usepackage{epsfig}
\usepackage{graphicx}
\usepackage{epstopdf}
\newcommand\ocen{$\omega$~Cen~}

\newcommand\mg{{M_{\rm g}}}
\newcommand\msun{\rm{M_{\odot}}}

\newcommand\teff{T$_{\rm eff}$}
\newcommand\tbce{T$_{\rm bce}$}
\def\simgt{\lower.5ex\hbox{$\; \buildrel > \over \sim \;$}}
\def\simlt{\lower.5ex\hbox{$\; \buildrel < \over \sim \;$}}
\def\c12{$^{12}$C}
\def\neo20{$^{20}$Ne}
\def\al27{$^{27}$Al}
\def\ne22{$^{22}$Ne}
\def\na23{$^{23}$Na}
\def\mg25{$^{25}$Mg}
\def\mag26{$^{26}$Mg}
\def\nafe{[Na/Fe]}

\def\ofe{[O/Fe]}

\def\alfe{[Al/Fe]}
\def\mgfe{[Mg/Fe]}

\title[Evolution of abundance patterns in Globular Clusters]
{Abundance patterns of multiple populations in Globular Clusters: a chemical evolution model based on yields from AGB ejecta}
\author[A. D'Ercole et al.]
{Annibale D'Ercole,$^1$\thanks{E-mail: annibale.dercole@oabo.inaf.it}  
Francesca D'Antona,$^2$ Paolo Ventura,$^2$
\newauthor
Enrico Vesperini$^3$ and Stephen L. W. McMillan$^3$\\
$^{1}$INAF- Osservatorio Astronomico di Bologna, via Ranzani 1, I-40127 BOLOGNA (Italy)\\
$^{2}$INAF- Osservatorio Astronomico di Roma, via di Frascati 33, I-00040 Monteporzio (Italy)\\
$^3$Department of Physics, Drexel University, Philadelphia, PA 19104, USA
}
\begin{document}
\def\enr#1{{\bf[#1 -- Enrico]}}
\def\steve#1{{\bf[#1 -- Steve]}}
\def\drex#1{{\bf[#1 -- Drexel]}}
\def\franca#1{{\bf[#1 -- Franca]}}
\def\ann#1{{\bf[#1 -- Annibale]}}

\date{Accepted ... Received ...; in original form ...}

\pagerange{\pageref{firstpage}--\pageref{lastpage}} \pubyear{2002}

\maketitle

\label{firstpage}

\begin{abstract}

  A large number of spectroscopic studies have provided evidence of
  the presence of multiple populations in globular clusters by
  revealing patterns in the stellar chemical abundances.  This paper
  is aimed at studying the origin of these abundance patterns.  We
  explore a model in which second generation (SG) stars form out of a
  mix of pristine gas and ejecta of the first generation of asymptotic
  giant branch stars.  We first study the constraints imposed by the
  spectroscopic data of SG stars in globular clusters on the chemical
  properties of the asymptotic and super asymptotic giant branch
  ejecta.  With a simple one-zone chemical model, we then explore the
  formation of the SG population abundance patterns focussing our
  attention on the Na-O, Al-Mg anticorrelations and on the helium
  distribution function.

  We carry out a survey of models and explore the dependence of the
  final SG chemical properties on the key parameters affecting the gas
  dynamics and the SG formation process. Finally, we use our chemical
  evolution framework to build specific models for NGC 2808 and M4,
  two Galactic globular clusters which show different patterns in the
  Na--O and Mg--Al anticorrelation and have different helium
  distributions.  We find that the amount of pristine gas involved in
  the formation of SG stars is a key parameter to fit the observed
  O--Na and Mg--Al patterns. The helium distribution function for
  these models is in general good agreement with the observed one. Our
  models, by shedding light on the role of different parameters and
  their interplay in determining the final SG chemical properties,
  illustrate the basic ingredients, constraints and problems
  encountered in this self-enrichment scenario which must be addressed
  by more sophisticated chemical and hydrodynamic simulations.
\end{abstract}


\section{Introduction} 
\label{sec:introduction}
In the last decade, a large number of spectroscopic and photometric
studies have provided strong indication of the presence of multiple
stellar populations in globular clusters. 

The spectroscopic evidence comes from the observed spreads and
anticorrelations between the abundances of light elements, not shown
by the halo stars of similar metallicities. Specifically, all the
clusters studied have shown an anticorrelation between Na and O and an
anticorrelation between Mg and Al has been found in several massive
clusters \citep[cf.][]{grasneca04}. The fact that these chemical
anomalies have been found also in turnoff and subgiant stars
\citep{gratton2001, ramirezcohen2002} indicates that they must be
already present in the gas from which these stars formed. The
anticorrelation between CN and CH bands observed in giant
\citep[e.g.][]{norrisfd1984} and turnoff stars indicates a difference
in the nitrogen abundance and is considered another fingerprint of the
presence of multiple populations in globular clusters.

An additional important constraint on the possible sources of gas from
which second generation stars formed is provided by the observed
constancy of the total C+N+O. The only exception, so far, is NGC 1851
for which a difference of ~0.6 dex is found among four red giants
\citep{yong2009} and for which the analysis of the subgiant branch
suggests that the total C+N+O increases by about a factor three
between the stars in the brighter and those in the dimmer branch
\citep{ventura18512009}.

Photometric studies have also provided a number of important results
indicating the presence of significant differences in the helium
abundances of stars within individual clusters. Indeed, the presence
of helium differences seems to be the only explanation for the
multiple main sequences or spread in the main sequence (MS) observed in
some clusters. These observations confirmed the prediction of the
presence of a helium spread based on the complex morphology of the
horizontal branches of some clusters made by \citep{daca04,dc2008}. In
extreme cases (the blue MS of \ocen\ and of NGC 2808) a helium content
Y=0.38--0.40 is inferred \citep{norris2004,dantona2005,pio07}.

A recent spectroscopic survey by \citet{carretta2009a} has  found
that all the 15 clusters studied show the spectroscopic evidence of the
presence  of multiple populations and that in all cases second
generation stars represent a significant fraction  (50-80 \%) of a
cluster population. This implies that matter forming more than half of
each cluster stars has been processed through the hot CNO cycle and by
other proton-capture reactions on light nuclei.

In spite of the light elements variation, most GCs are mono--metallic
objects, as far as abundances of heavier elements are concerned
\citep[see][for a recent review on this subject]{grasneca04}. Their
heavy (Z $>$13) element metallicity, usually represented by the ratio
[Fe/H], is found to be extremely homogeneous from star to star in each
cluster. \cite{carretta2009c} find that the upper limit to the scatter
of iron is less than 0.05~dex in the 19 clusters they examine. This is
consistent with previous determinations, e.g. for NGC~6752
\citep[$\sigma$=0.02][]{yong2005}, and for the clusters examined by
\cite{kraftivans2003} (0.03 -- 0.10~dex). The width in color of the
main sequence and/or of the red giant branch also agree with the
spectroscopic determination \citep{suntzeff1993}.  As the site of
production of heavy elements ($\alpha$--capture and Fe--group
elements) are stars exploding as core-collapse or thermonuclear
supernovae \citep[e.g.][]{wheeler1989}, the star to star iron
homogeneity means that supernova ejecta do not affect the chemistry of
the gas processed through the hot CNO cycle. On the other hand,
significant iron and s--process elements spread is well known to be
present in \ocen\ \citep[e.g., for a recent analysis,][and references
therein]{johnson2009} and more recently it has been confirmed in other
massive clusters like M22 \citep[][and references therein]{Marino09,
  dacosta2009}, M54 \citep[][and references therein]{sl1995,
  bellazzini2008}, and Terzan~5 \citep{ferraro2009}. In smaller
clusters like NGC~1851, the SG might be enriched in calcium, according
to \cite{Lee09}, but see the discussion by \cite{carretta2010}.

In \citet{der08} (hereafter Paper I), we presented a model for the
formation and dynamical evolution of multiple populations in globular
clusters (GCs). In particular, we explored a model in which second
generation (hereafter SG) stars form out of the ejecta of the first
generation (hereafter FG) stars. By means of hydrodynamic
simulations we have shown that the ejecta of AGB (asymptotic giant
branch) stars collect in a cooling flow into the cluster core,
where they form a subsystem of SG stars initially strongly
concentrated in the cluster innermost regions. By means of N-body
simulations we have then explored the subsequent stellar dynamical
evolution of the cluster focusing our attention on the early loss of
FG stars, on the evolution of the ratio of the number of FG and SG
stars and on the evolution of their relative spatial distribution and
mixing.

Paper I also included a preliminary discussion on the chemical
abundances of FG AGB stars and on the possibility of reproducing the
helium excess and other chemical anomalies observed in SG stars.
Several theoretical and observational papers have shown that dilution
of the FG ejecta with gas having the pristine composition is needed to
explain the general shape of the anticorrelations \citep{prantzos2007,
  bekki07,dantonaventura2007,carretta2009a}. If the SG anomalies are
inputed to processing by Hot Bottom Burning (HBB) at the bottom of the
convective envelope of massive AGBs, these models produce a direct
correlation between sodium and oxygen abundance in the processed
matter. This direct correlation is unavoidable, as found in all the
relevant AGB computations
\citep{dh2003,herwig04,karakas2007,ventura2008a} and as explained in
more detail in Section 2. While the total sodium yield is dependent on
many uncertain factors, it is certain that a small oxygen depletion
(lower burning temperature) goes together with a larger sodium
abundance and a large oxygen depletion (higher burning temperature) is
accompanied by a smaller sodium abundance. Any attempt to reverse this
direct correlation, in the limits provided by available cross
sections, has been unfruitful \citep[e.g.][]{ventura2008a}. An
observational hint in favour of the above arguments is given by
\citet{carretta2009a} who find a direct correlation between the
minimum O and the maximum Na abundances of the 15 clusters they
studied; such a result could not be understood if O and Na of the AGB
ejecta were anticorrelated.

From all the above, it is apparent that, in order to explain the GC
chemical patterns, it is necessary to introduce a form of dilution of
the AGB ejecta, either with pristine gas, or with gas not showing the
peculiar abundance patterns of the hot--CNO processed matter.

The source of this pristine matter still requires a detailed
understanding.  \cite{der08} proposed that initial asymmetries in the
gas distribution allow to vent out the Type II supernovae (SNe~II)
ejecta along preferential directions, creating an ``hour-- glass"
cavity and leaving some pristine gas in a torus at the outskirts of
the cluster; at the end of the SN II activity this torus eventually
collapses back into the cluster. In this scenario, the torus is not
affected by contamination from supernova ejecta, and the iron content
of the diluting matter is still the pristine one, consistently with
the absence of iron differences between the FG and SG summarized
above.  While the evolution of the larger clusters is probably also
affected by more prolonged star formation and supernovae type Ia
contamination \citep[e.g.][]{mar07}, the presence of slight metal
increase in smaller clusters might be a possible indication that the
``pristine" gas in the falling back torus has been partially
contaminated by the SN~II ejecta.

This paper is aimed at significantly expanding the initial study of
SG stars chemical abundances presented in Paper I and at exploring in
detail the SG chemical anomalies emerging from a model in which SG
stars form from a mix of AGB ejecta and gas with pristine chemical
composition. 

We will refer to the scenario presented in Paper I.
Consequently, we need to associate to the temporal
evolution of the FG clusters the chemical composition of the ejecta of
super--AGB and then massive AGB stars (hereinafter simply referred to
as AGB pollutors) that successively provide the matter used for
building up second generation stars. The chemical composition of the
H-rich matter in the envelopes of these stars is affected by the hot
CNO cycle processing at the basis of the convective envelope of these
giants (hot bottom burning, hereafter HBB) and possibly also by the
third dredge up (3DU) occurring after each thermal pulse. Stellar
winds and planetary nebula ejection deposit the matter of these
envelopes into the interstellar medium (ISM), where it can contribute to
the formation of the second generation. 

Despite the significant effort devoted to the determination of the
yields of the massive AGBs and the numerous attempts to quantify the
uncertainties in these calculations
\citep{ventura2005a,ventura2005b,karakas2007}, there are still large
differences in the results obtained by different groups and even by
the same groups at different times and no general consensus has been
reached.

Our aim is to focus on the chemical properties of the stellar
envelopes and explore the constraints on these
properties imposed by the spectroscopic data of SG stars in globular
clusters. We have addressed this problem by means of a simple one-zone
model in which the ISM is supplied by the FG AGB ejecta and by the
possible accretion of pristine gas, and is depleted by the SG star
formation (hereafter SF).  This simplified approach allows to easily
control the evolution of the amount of the AGB ejecta and pristine gas
involved in the SG formation process and to fully explore the
chemical properties of the ISM and of the SG stars.

The structure of the paper is the following. We start in sect.
\ref{sec:yelds} with an overview of the chemical abundances of the
massive AGB ejecta. We then illustrate our general assumptions in
sect. \ref{sec:assump} and the model details in sect.
\ref{sec:comp}. We then analyze the model in sect. \ref{sec:moana},
and compare our results with the data of NGC 2808 and M4 in
sect. \ref{sec:ngc2808} and sect. \ref{sec:m4}, respectively. Finally,
we summarize our conclusions in sect. \ref{sec:conc}.

\section{Educated guesses on the yields from super--AGB and massive AGB stars}
\label{sec:yelds}

\begin{figure}    
\centering{
\includegraphics[width=8cm]{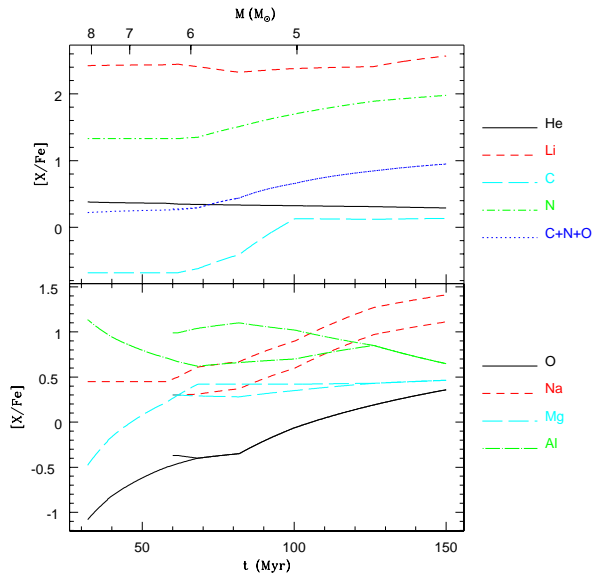}
}
\caption{Time evolution of the chemical abundance of the ejecta of the
  dying stars adopted in this paper (thick lines) and as given by
  \citet{ventura2009} (thin lines). On the top axis some masses of
  dying stars are reported in correspondence to their lifetime. }
\label{fig:tab} 
\end{figure}

We start by re--examining all the uncertainties connected
with the computation of the chemical yields of the AGB pollutors,
identifying the range of possible models producing chemical patterns
and trends consistent with those found in spectroscopic observational
studies.  We will adopt yields for massive AGBs listed in Table 2 in
\citet{ventura2009} for metallicity Z=10$^{-3}$ (see
Fig. \ref{fig:tab}), and in Table \ref{chem}, modified and
extrapolated as shown in the columns labelled ``n". For carbon and
nitrogen, the abundances of \citet{ventura2009} are adopted without
modification (for 6.5 and 9 $\msun$, the values of M=6.3 $\msun$ have
been adopted).

\begin{table*}
\caption{Averaged abundances in the ejecta of massive AGB and super--AGB stars.}             
\label{chem}      
\begin{tabular}{c c  c c c c c c c c c c c c}     
\hline       
$M/M_{\odot}$ & $\tau/10^6$$^a$ & $M_{\rm c}/M_{\odot}$$^b$ &  Y& \ofe$^d$ & \ofe$_n^e$  & \nafe$^d$ & \nafe$_n^e$ & \mgfe$^d$ & \mgfe$_n^e$ &\alfe$^d$ & \alfe$_n^e$ \\
\hline                                                                            
   3.0 & 332   & 0.76 &  0.248 &  0.92 &  0.92 & 1.16 & 1.46  & 0.57  & 0.57  & 0.65 & 0.65   \\  
   3.5 & 229   & 0.80 &  0.265 &  0.77 &  0.77 & 1.30 & 1.60  & 0.55  & 0.55  & 0.66 & 0.66   \\
   4.0 & 169.5 & 0.83 &  0.281 &  0.44 &  0.44 & 1.18 & 1.48  & 0.48  & 0.48  & 0.55 & 0.55   \\
   4.5 & 130.3 & 0.86 &  0.310 &  0.19 &  0.19 & 0.97 & 1.27  & 0.43  & 0.43  & 0.85 & 0.85   \\
   5.0 & 103.8 & 0.89 &  0.324 & -0.06 & -0.06 & 0.60 & 0.90  & 0.35  & 0.42  & 1.02 & 0.70  \\
   5.5 & 85.1  & 0.94 &  0.334 & -0.35 & -0.35 & 0.37 & 0.67  & 0.28  & 0.42  & 1.10 & 0.66   \\
   6.0 & 71.2  & 1.00 &  0.343 & -0.40 & -0.40 & 0.31 & 0.61  & 0.29  & 0.42  & 1.04 & 0.62   \\
   6.5 & 61.5  & 1.05 &  0.360$^c$  &          & -0.50 &         & 0.45  &          & 0.22  &         & 0.71   \\
   9.0 & 32.0  & 1.37 &  0.380$^c$  &          & -1.10 &         & 0.45  &          & -0.50 &         & 1.15   \\
\hline                                                                                                       
\end{tabular}
\leftline{$^a$Total evolutionary time until the AGB phase }
\leftline{$^b$Core mass at the beginning of the AGB phase.}
\leftline{$^c$From \citet{siess2007}.}
\leftline{$^d$From \citet{ventura2009}.}
\leftline{$^e$Values adopted in this paper.}
\end{table*}

 
We list in the next subsections the problems relative to each
individual chemical element, and how they are related to the physical
inputs (e.g. cross sections) or to the treatment of some important
structural problems (e.g.  the treatment of convection, mass loss,
possible extra--mixing at the formal convective boundaries).  We will
mainly deal with the yields relative to the metallicity Z=10$^{-3}$,
[$\alpha$/Fe]=0.4, for two reasons: 1) this metallicity is adequate to
describe most of the GCs for which data relevant for the study of the
chemical anomalies are available; 2) this is the metallicity for which
most theoretical models have been built \citep[see][for yields
calculated for different values of Z]{ventura2008b, ventura2008a,
  ventura2009}.

\subsection{The role of super--AGBs and their yields}
\label{subsec:supagb}

Most researchers focused on the problem of multiple stellar
populations mainly when two clusters (\ocen\ and NGC~2808) were
recognized to have a large fraction of main sequence stars
($\sim$25\% in \ocen, $\sim$15\% in NGC~2808) in a ``blue MS'' which
could only be interpreted as a very helium rich MS \citep{bed04,
  norris2004, piotto2005, dantona2005, pio07} \footnote{Actually,
  \cite{pio07} show that in NGC~2808 three different parts of the MS
  can be identified.}.  In both clusters, this extreme--helium
population must also be very {\it homogeneous}, both in helium and
metal content, otherwise the blue MS would not be detached from the
rest of MS stars.  Two other very massive clusters, NGC~6441 and
NGC~6388, show an horizontal branch (HB) extended to very high-\teff,
and their RR Lyr variables have very long periods, indicating very
high RR Lyr luminosity, a peculiar occurrence at the high metallicity
of these clusters \citep{rich1997}.  Also in these cases, a very high
helium population (including $\sim 10$\% and $\sim 20$\% of stars
respectively) provides the only reliable explanation for these
features \citep{caloi2007, busso2007,yoon2008},

In all cases, modelling the MS colours, or the HB morphology and the RR
Lyr periods, the helium content of the most extreme population results
to be Y$\sim$0.38--0.42. Values so high are extremely peculiar,
particularly because the helium enrichment {\it is not} accompanied by
metal enrichment (a value $\sim 70$\ is estimated for the parameter
$\Delta Y/ \Delta Z$). Models of massive AGBs can reach Y values of
$\sim 0.35$\ in their envelopes, and therefore models based on the
formation of this high helium population out of massive AGBs ejecta
can not reproduce the high Y values suggested by observations
\citep{karakas2006}. Similarly, models based on yields coming from
either Type II or Type Ia supernovae (SNe Ia) are not successful
\citep{romano2007, mar07,choi2008}.  \cite{pumo2008} noticed that a
very high helium content (Y up to $\sim$0.38) can be achieved already
just after the second dredge--up in the envelopes of the most massive
super--AGB' stars, as shown by the models by \cite{siess2007}.
Super--AGBs are the stars that reach a carbon oxygen core mass of
$\sim 1.05$ $\msun$, and ignite carbon in conditions of
semi--degeneracy. The final outcome of this burning is a degenerate
oxygen--neon core, that can now traverse a "normal" thermal pulse (TP)
phase, like the stars with C--O core.  A successful preliminary
chemo--dynamical model, able to explain the helium distribution of the
triple main sequence of NGC~2808, was then proposed in Paper I. We
assumed that a ``first" SG formed out of the ejecta of super--AGBs, so
that this population has a very high Y, similar for all stars, while,
as the SG formation phase proceeds, a ``second" SG forms out of gas
from massive AGBs mixed with pristine gas falling back onto the
central regions and has a smaller helium content.  This model produces
a gap in the helium distribution of SG stars and can successfully
reproduce the distinct populations identified in the color-magnitude
diagram of NGC~2808.

To make further progress in understanding the origin of SG stars and
to test the viability of different hypotheses on the origin of the gas out of
which SG stars formed, it is essential to explore whether all the
chemical anomalies that characterize SG stars can be successfully
reproduced.  Unfortunately, only a few models of super--AGBs
are available in the literature, and no low metallicity mass has been
evolved through the whole TP phase. For the case of helium, we can
rely on the helium abundance reached in the envelope after the second
dredge up, as computed by \cite{siess2007}, but we do not know how
precisely the other abundances will be modified by HBB during the rest
of the stellar life, until the whole envelope is lost to the
interstellar medium.  In addition, even when the first models through the super--AGB TP
phase will be available, their input physics must be carefully explored,
as done for the massive AGB stars. Until then, we must make guesses on their
yields.  In this paper, we simply propose that the yields from
super--AGBs for O, Na, Mg, Al, are compatible with the patterns
observed in the most extremely anomalous GC populations. For instance,
the sum of the CNO abundances can remain close to the initial value
during the thermal pulse phase of super--AGBs, as they have a limited
efficiency of third dredge up, because of the weak helium luminosity
reached during the thermal pulses \citep{siess2007b}. We then {\it
  extrapolate} the yields computed for the lower AGB masses to the
super--AGB regime.  The extrapolation must satisfy two constrains: 1)
super--AGBs have oxygen--neon core masses larger than the largest
carbon--oxygen core masses of the most massive AGBs that do not ignite
carbon. Consequently, we expect more extreme physical conditions in
their convective envelopes, than in those of the most massive AGBs we
have computed, and a more advanced nucleosynthesis due to
proton-capture elements (this assumption also depends on the effect of
mass loss, and on the total duration of the thermal pulse phase); 2)
in addition, the results must not be in deep contrast with the
observed abundance patterns. As for the effect of the third dredge up,
in this work we will assume that dredge up in super--AGBs, if any, is
less efficient than in less massive AGBs, as suggested by
\citet{siess2007b}.

\subsection{Oxygen burning}
\label{subsec:ona}
One of the most challenging issues in modelling the abundance patterns
of SG stars in globular clusters is the SG low oxygen  abundance. 

Population II stars are in general $\alpha$--enriched, and thus
oxygen--enriched, due to the role of SNII in the oxygen production
\citep[e.g.][]{matteucci1986}; in order to deplete the abundance of
oxygen, the gas from which SG stars are formed must have gone through
the full CNO cycle (the ON chain) that occurs at temperatures of $\sim
40$ MK in the stellar interior, and needs temperature of \tbce $\sim
60-80$ MK at the bottom of AGB convective envelopes (\tbce), both due
to the much lower stellar densities of these regions, and to the
shorter evolutionary times of this phase \citep[see,
e.g.][]{prantzos2007}.

Efficient ON burning occurs due to HBB in massive AGBs ---and
consequently in super--AGBs, due to the higher temperatures reached at
the bottom of their convective envelopes.  \cite{ventura2005a} showed
that more efficient ON processing is favoured by more efficient
convection.  However, the most massive AGBs with values of the
metallicity (Z$\sim 10^{-3}$) typical of the clusters showing very
strong chemical anomalies, can not deplete oxygen by more than $\sim
0.8$ dex.  Starting form a typical [O/Fe]=+0.4, the minimum oxygen
abundance is [O/Fe] $\sim$ --0.35 -- --0.40.  In most GCs the
O--depletion is limited to $\simlt$0.8 dex, but we do know that giants
in clusters like M~13 \citep{sneden2004} or NGC ~2808 \citep{car08}
show extreme stars with oxygen depletion down to a factor larger than
20.

There are two options to explain the oxygen abundance of these extreme
stars: 1) either the strong O-depletion is due to the higher \tbce\
reached by super--AGBs; 2) or it is a consequence of in--situ mixing
in red giants (these extreme anomalies have been so far found only
along the giant branch) suffering deep extra--mixing due to their very
high initial helium content. Indeed, in these stars, the molecular
weight barrier that could forbid deep mixing at luminosities below the
red giant bump is much smaller than in stars with normal Y
\citep{dantonaventura2007}.

In both cases, if we wish to have an educated guess about the oxygen
yield of super--AGBs, the choice is mandatory: it must be {\it
  smaller} than for the maximum computed AGB mass, so we adopt
[O/Fe]=--1.1 for the progenitor largest mass 9 $\msun$.

\subsection{Sodium production}
Understanding the sodium abundance in the SG stars presents similar
complexities. The sodium yield is provided by three different stages
of the life of a massive AGB star. At each of these stages, different
problems must be attacked to determine the yields, and may have
different solutions.  In addition, during these stages an AGB star
loses mass, so the total yield will also depend on the mass loss rate
in each of these phases.  We have discussed these problems in a series
of papers dedicated to the nucleosynthesis in AGBs \citep[][and
references therein]{ventura2008b}. Here below we outline the phases
during which the sodium changes:
\begin{enumerate}
\renewcommand{\theenumi}{(\arabic{enumi})}
\item The first phase during which sodium increases is when the second
  dredge up, before the star climbs up the AGB, mixes to the surface
  the \na23 and \ne22 produced inside the star during the previous
  H--burning phases.
\item Immediately after, the reaction \ne22(p,$\gamma$)\na23 takes
  place, due to the HBB, and \na23 begins to be cycled back through
  the reaction \na23(p,$\alpha$)\neo20.  The most important of these
  two paths in order to determine the total \na23 yield is the first
  one, \ne22(p,$\gamma$)\na23, that, unfortunately, is uncertain {\it
    by a factor $\sim$2000} in the range of temperature of interest
  for oxygen depletion \citep{hale2004}.  It is important to point
  out, however, that, once all the \ne22 is burned to \na23 via HBB,
  the sodium production does not increase even if we further increase
  the cross section of this reaction \citep{izzard2007}. We can keep
  more sodium only by lowering the reaction rate
  \na23(p,$\alpha$)\neo20 \citep{ventura2006}.
\item The third stage by which sodium in the envelope changes is the
  third dredge up phase (3DU) following each thermal pulse. The
  smaller the initial mass, the more effective the 3DU becomes. At
  each episode, the 3DU in fact mixes down into the helium layers the
  nitrogen obtained in the H--envelope due to HBB--CN cycling of the
  \c12 produced during the thermal pulse and partially dredged
  up. Consequently, {\it primary} \ne22, produced by the chain
  $^{14}$N($\alpha,\gamma$)$^{18}$F($\beta^+,\nu$)$^{18}$O($\alpha,\gamma$)\ne22,
  is also dredged up from the helium inter--shell, and is converted
  into primary sodium by HBB.  If the stars go through many episodes
  of 3DU, this third phase can increase the sodium abundance even by
  orders of magnitude, but it also leads to a strong increase in the
  total C+N+O yield (see sect. \ref{subsec:cno}) which has not been
  observed in globular clusters.
\end{enumerate}

All the computations of AGB evolution in the literature clearly show
that, as the AGB mass increases, the sodium yield decreases; this is a
consequence of the increasing \tbce.  A larger \tbce\ depletes more
oxygen, and destroys sodium more easily by the \na23(p,$\alpha$)\neo20
cycle. In addition, in smaller masses the 3DU will dredge up more the
\ne22 and convert it into sodium. This implies that the massive AGB
models, and probably the more massive super--AGBs as
well, provide a {\it direct} O--Na correlation, and not the observed
anti--correlation.  We address this problem in sect. \ref{subsec:rm}
and show the possible role played by dilution with pristine gas in
turning this correlation into the observed Na-O anti-correlation.

In order to reproduce the observed abundances of Na in the SG, we need
to have {\it the largest possible sodium yields for the masses in
  which the 3DU does not play a strong role}, that are, as we will see
from the C+N+O examination, M$\simgt 5$ $\msun$.  Starting from the reference
values of the yields from the calculation of \cite{ventura2009} (see
their Table 2 and Table 1), there are two key ingredients which are
still affected by significant uncertainties and could lead to larger
sodium abundance in the massive AGB envelopes:
\begin{enumerate}
\item{ {\bf the \neo20(p,$\gamma$) reaction rate:} 
 a 50 \% increase in the \neo20(p,$\gamma$)
 rate is  well within the NACRE \citep{angulo} reaction rate
 uncertainties and, keeping all other parameters 
 fixed, will lead to a 0.15 dex increase in the sodium yield,
 because the equilibrium of the Ne--Na chain is shifted towards \na23;}
\item{{\bf  the \neo20 abundance:}
the initial \neo20 abundance is not well known\footnote{
In recent years, an increase in the \neo20 abundance for the Sun
has been suggested, as it could produce the opacity increase necessary to
obtain a better fit of the solar oscillation spectrum, once the iron and
oxygen content are reduced following the results of the 3D analysis
of the solar spectrum \citep[for a summary of this complex problem, see][]{asplund2009}.},
and it may also vary from cluster to cluster (sect. \ref{sec:comparison}). 
In our reference computation, the \neo20 abundance is already 
$\alpha$--enhanced, as for oxygen and the other $\alpha$-elements,
by 0.4 dex with respect to the solar ratios. If we
increase this initial \neo20 by another factor 2,
 the final [Na/Fe] is increased by $\sim$0.2 dex (for the important
 masses from 6.3 to 5 $\msun$). Owing to our numerical tests,
in a first approximation we can assume:
 \begin{equation}
     \delta [Na/Fe] \simeq \delta [ ^{20}Ne/Fe]
\label{eqna}
 \end{equation}    
 }
\end{enumerate}

If we adopt both the enhanced reaction rate and the larger \neo20
abundance, the combined effect of these two assumptions leads to
[Na/Fe]=[Na/Fe]$_{\rm standard}$+0.35.  In our calculations the
increase of the sodium yield is calibrated by comparison with the
observational data by +0.3dex (see Table \ref{chem} and
Fig. \ref{fig:tab}) for the masses up to 6 $\msun$.  For the
super--AGB stars, we decrease the sodium content to [Na/Fe]=0.45,
based on the hypothesis that these stars will be more efficient in
destroying sodium.  We warn that this suggestion will need
confirmation from real computation of super--AGB evolution.  Once we
have assumed a "standard" table of abundances with increased sodium,
we still can use equation \ref{eqna} to model clusters with abnormally
larger $\alpha$-elements abundances (see the case of M4 in
sect. \ref{sec:m4})

\subsection{The total CNO abundance in the SG}
\label{subsec:cno} 
For C, N and O we will use the yields from Table 2 of
\cite{ventura2009} without any adjustment. For the masses from 6.5 to
9 $\msun$, the super--AGB regime, we adopt the C and N envelope
abundances of the most massive computed AGB model of 6.3 $\msun$. The O
abundance in this mass range, instead, has been calibrated to fit the
abundance of the O-poor stars present in NGC 2808, as discussed in the
previous subsection (cf. Fig. \ref{fig:tab}).

As most of the SG stars in GCs do not show a significant increase of the
total CNO \citep{ivans1999, ivans2001, carretta2005}, the matter out
of which the SG forms can not come from stars in which the effect of
the 3DU is relevant. A reasonable assumption based on the data is that
the total CNO in most clusters so far examined is within a factor two
of the CNO of the FG.  If we adopt the yields from 
\cite{ventura2009} as our reference values, and assume that SG stars
form only from pure AGB ejecta, the limiting mass for the SG formation
is then $\sim 5$ $\msun$, and the limiting age $\sim$100 Myr (time
evolution of the 5 $\msun$, see Table \ref{chem}). If, on the
other hand, the matter from which SG stars form is diluted with
pristine gas, as shown to be necessary in sect. \ref{sec:moana}, the
limiting age for SG formation can be a bit larger.

An interesting indication concerning possible variations in CNO comes
from recent spectroscopic observations of a few red giants in
NGC~1851. These observations have shown that the C+N+O abundance of
the stars observed varies by up to a factor 4 \citep[0.6 dex,
][]{yong2009}.  In addition, new HST photometric data
\citep{milone2008} have revealed a split subgiant branch. Two
interpretations are possible: either the stars in the lower subgiant
branch are older by $\sim 1$Gyr \citep{milone2008}, or they are SG
stars with a larger total C+N+O content \citep{cassisi2008}.
\cite{ventura18512009} show that the total C+N+O must be about a
factor three larger in the lower subgiant branch stars. They argue
that the matter from which the SG formed should have been diluted at
50\% with pristine matter having approximately the composition of the
ejecta of $\sim 4$ $\msun$\ AGBs. With this choice, they can explain at
the same time the C+N+O abundance and the lack of a significant helium
enrichment, that would be recognized in the luminosity of the blue
horizontal branch stars \citep{salaris2008}.  In our scenario, the SG
in this case should have formed from matter collected in the cluster
until an age of $\sim 150$ Myr, at the evolution of the 4 $\msun$
\citep{ventura18512009}.

In the construction of possible models for the SG, then, the C, N and
O composition of the SG is an important parameter, and it is also
important to obtain further observations of its range of variability
for as many clusters as possible. We point out that the AGB models of
\cite{ventura2009} adopted here have smaller CNO enhancement than
other published models \citep[e.g.][]{karakas2007}, as they have a
smaller number of TPs and 3DU episodes. This is due both to the mass
loss rate adopted and to the use of an efficient convection model that
provides a larger stellar luminosity (and consequently a larger mass
loss rate) for a fixed core mass.  Should we adopt models with a more
efficient 3DU, or a larger number of TPs, the formation of the SG must
occur at shorter age (larger evolving initial masses) both for the
''normal" SG clusters (no CNO enrichment) and for clusters like
NGC~1851, where the CNO enrichment is present.

\subsection{The Magnesium and Aluminium yields}
\label{subsec:mgal}

In some (massive) GCs only, also an anticorrelation between magnesium
and aluminium is revealed. The depletion in total Mg is restricted to
0.2 -- 0.4~dex.  Generally, the stars with the highest degree of
oxygen depletion are also strongly enriched in Al, up to the values
[Al/Fe]$\sim 1 - 1.2$ found in the red giants of M13 and M15
\citep[for a summary of the data of the last decade, see
][]{ivans1999}.  The anticorrelation shows that the matter of the SG
has been subject to the Mg--Al chain, that, favouring proton captures
by the heavy isotopes of magnesium, eventually leads to the synthesis
of \al27\ \citep{dh2003,ventura2005a}. In models with many third
dredge up episodes, penetration of the convective envelope into the
layers touched by 3$\alpha$\ burning may bring to the surface \mg25\
and \mag26\ synthesized via $\alpha$ captures on \ne22\ nuclei. These
isotopes produce \al27\ by proton capture, once the H--shell is
reactivated \citep{ventura2008a}.  The recent analysis of Mg and Al
abundances from UVES spectra in 18 GCs, by \cite{carretta2009b}, shows
however large cluster to cluster variations in the Al--Mg
anticorrelation, and that in most cases the Mg variation is very
limited, even for stars with [Al/Fe]$\sim $0.8.  They also show that the
maximum Mg abundance and the minimum Al abundance (that should be
indicative of the composition of the FG) may be significantly
different in different clusters --even by an order of magnitude for
Al. Where the Mg abundance is larger than the typical [Mg/Fe]+0.4
assumed for the $\alpha$--enhanced composition, this may reflect
different overabundances in the matter forming the FG, and we must
take it into account, both for magnesium and for the other $\alpha$
elements (see sect. \ref{sec:m4}).

In their theoretical study, \cite{ventura2008a, ventura2009} adopted
the NACRE upper limits for the aluminium production from proton
captures on $^{25}$Mg and $^{26}$Mg, in order to obtain very
significant Al production in their massive AGB models. Their aim was
to reproduce the extended Mg--Al anticorrelation present in the M13 and
M15 red giants.  Their results, however, are in contrast with the
abundance patterns in M4 (NGC~6121) in which the maximum [Al/Fe] is
$\sim$0.8, and the {\it minimum} is [Al/Fe]$\sim 0.6$
\citep{ivans1999, marino2008}, so that this cluster does not show
great Al variations, if any.

As the large Al abundances of NGC 2808 seem to refer only to very
oxygen poor stars, we can attribute them to the super--AGB evolution.
Consequently, we decided to restore the above cross section to their
standard values, and assume that significant Al production is limited
to the super--AGB masses only. For the temperatures of interest here
($T \sim 10^8$K), the NACRE recommended values for the
$^{25}$Mg(p,$\gamma)^{26}$Al and $^{26}$Mg(p,$\gamma)^{27}$Al
reactions are smaller, respectively, by a factor of 2 and 4 than the
rates adopted by \cite{ventura2009}.  The aluminium production is
consequently reduced: its average content in the ejecta of massive
AGBs of 5 and 6$M_{\odot}$ with metallicity Z=0.001 diminishes to
[Al/Fe]=0.7 and 0.62 when the recommended cross sections for the
proton capture reactions by magnesium isotopes are adopted, to be
compared to the published values of [Al/Fe]=1.02 and 1.04.  The
modifications to the Mg and Al abundances are provided in Table \ref{chem}.

\section{General assumptions of the model}
\label{sec:assump}
The general assumptions driving the present models are the same
adopted in Paper I, and we recall them here briefly. We assume that
the FG stars are already in place and have the same chemical
abundances of the pristine gas from which they form. The SG stars form
from AGB ejecta. In order to have a current similar number of SG and
FG stars, the FG population must have been initially about ten times
more massive so as to supply enough mass in AGB ejecta to form a
sizable SG population. This conclusion remains substantially valid
even if, during its evolution, the GC accretes some external pristine
gas. A very large initial FG is invoked also by \citet{ren08} on the
basis of the large amount of helium observed in some multiple
populations of very massive GCs (see sect. 2.1).

We assume that only SG stars with $M\leq 9$ $\msun$ can form
(cf. Paper I). More massive stars explode as SNe II, but the very
small spread of iron in GCs \citep[c.f.][]{carretta2009c} indicates
that the SG stars hardly have been polluted by SNe II (we discuss this
point in the Introduction and in sect. \ref{sec:conc}). Moreover, as
shown in sect. \ref{sec:comparison}, the AGB pollution takes tens of
Myr to produce the observed chemical patterns; as the first SNe II
starts to explode after $\sim 5$ Myr, they would vent away all the
gas, preventing the formation of a substantial amount of SG stars
\footnote{On the contrary, massive stars in the FG must be present for
  two reasons. First, as the FG is initially nearly ten times larger
  than today (in order to deliver enough ejecta to form the observed
  amount of SG), SN explosions allow the cluster to get rid of the
  mass excess in a time shorter than the Hubble time (cf. Paper I);
  second, Extreme, very O-poor stars present in some cluster (e.g. NGC
  2808) can arise only from a pure AGB ejecta, and this could not be
  possible without the previous action of the FG SNe II clearing
  initially the GC of its residual pristine gas.}. Given a Kroupa
initial mass function (IMF) with an upper mass of 9 $\msun$, the mass
of the long-lived ($M\la 0.8$ $\msun$) SG stars is 54\% of the total
forming SG population.

\section{Computational method}
\label{sec:comp}

The computational framework adopted for this study is based on a
one-zone model for globular clusters in which the evolution of AGB
ejecta and pristine gas - from which SG stars form - is regulated by a few
parameters. The variation of these parameters can be easily
controlled and this allows us to explore their relative importance on
the resulting SG chemical properties.

The evolution of the gas and the process of SG formation is modelled by
the following equations which are integrated by means of a
fourth-order Runge-Kutta method:

\begin{equation}
\label{eq:den}
\dot \rho(t)=\alpha \rho_{\rm *,FG} +\dot \rho_{\rm pr}(t)-\nu {\rho(t) \over t_{\rm sf}} 
\end{equation}
\begin{equation}
\label{eq:elm}
\dot \rho^{\rm k}_{\rm ch}(t)=\alpha^{\rm k} \rho_{\rm *,FG}+\beta^{\rm k} \dot \rho_{\rm pr}(t)-\nu {\rho^{\rm k}_{\rm ch}(t) \over t_{\rm sf}} 
\end{equation}
\begin{equation}
\label{eq:pri}
\dot \rho_{\rm pr}(t)={\rho_{0,\rm pr}\over \sqrt{\pi} \tau}\exp{-[(t-t_{\rm ac})/\tau]^2}-\nu {\rho_{\rm pr}(t) \over t_{\rm sf}}
\end{equation}
\begin{equation}
\label{eq:str}
\dot \rho_{\rm *,SG}(t)=\nu {\rho(t) \over t_{\rm sf}}
\end{equation}

\noindent
Here below we discuss each of the above equations:
\begin{description}
\item {\it Eq. \ref{eq:den}} - This equation describes the time
  evolution of the total gas density $\rho$. The first term in the
  right hand side (RHS) represents the increment of $\rho$ due to the
  gas lost by the dying stars.  In principle, we would take into
  account the mass return from the FG stars as well as from the newly
  formed SG stars.  These latter, however, are much less than the
  former, and we neglect their contribution. We thus consider only the
  density $\rho_{\rm *,FG}$, and the specific mass return rate is
  given by $\alpha = {\dot \rho_{\rm *,FG}} /\rho_{\rm *,FG}$; this
  rate is computed following \citet{cioetal91} (see their eq. 3),
  adapted to the Kroupa IMF adopted here. The second term in the RHS
  of eq. \ref{eq:den} takes into account the gas increment due to the
  accretion of the pristine gas whose density is $\rho_{\rm
    pr}$. Finally, the last term describes the reduction of gas
  because of the star formation, and is characterized by the SF
  efficiency $\nu$ and the SF timescale $t_{\rm sf}$. The former is a
  free parameter summarizing our uncertainties about star formation
  and with values in the range $0<\nu <1$; the latter coincides with
  the free-fall time, and is thus proportional to
  ($\rho^{-1/2}$)\footnote{In a full hydrodynamic treatment, as in
    Paper I, $t_{\rm sf}$ is defined as the maximum between the local
    cooling time and the local free-fall time, but the first one is
    usually much shorter}.
\item {\it Eq. \ref{eq:elm}} - Besides the total gas density, it is
  necessary to evaluate the density of the single chemical elements we
  are interested in. In our models we take into account the abundance
  evolution of nine elements: He, Li, C, N, O, Na, Mg, Al,
  Fe. Equation \ref{eq:elm} represents a system of $kmax=9$ equations,
  each modelling the evolution of the density $\rho^{\rm k}_{\rm ch}$
  of the $k-th$ element. In these equations $\alpha (t) ^{\rm k}$ and
  $\beta ^{\rm k}$ represent the mass return rate of the $k-th$
  element in the FG, and the mass fraction of the $k-th$ element in
  the pristine gas, respectively.  The latter is determined by the
  assumed initial chemical composition of the pristine gas, while the
  values of $\alpha (t) ^{\rm k}$ are obtained at each time by the
  interpolation of the values given by our adopted ejecta abundances
  (cf. Fig. \ref{fig:tab} and Table \ref{chem}).
\item {\it Eq. \ref{eq:pri}} - This equation describes the evolution
  of the pristine gas density $\rho_{\rm pr}$. The two terms on the
  RHS model the time evolution of the pristine gas accretion (first
  term) and the SG star formation (second term). The gaussian form
  adopted to model the pristine gas accretion is arbitrary, but allows
  us to easily explore different accretion scenarios. The time $t_{\rm
    ac}$ regulates the time at which the maximum accretion rate
  occurs, and the time $\tau$ controls how the accretion evolves: a
  small value of $\tau$ corresponds to a sudden accretion, while a
  large value gives rise to a more gradual growth.  Finally, the
  constant $\rho_{0,\rm pr}$ controls the amount of available pristine
  gas.  The three constants $t_{\rm ac}$, $\tau$ and $\rho_{0,\rm pr}$
  must be regarded as free parameters of our models.
\item {\it Eq. \ref{eq:str}} - Finally, equation \ref{eq:str}
  regulates the growth of the SG stellar population.
\end{description}

In addition to $\rho_{0,\rm pr}$, $\nu$, $t_{\rm ac}$, $\tau$, one
more parameter must still be added: the time $t_{\rm end}$ at which
the simulations stop. It regulates not only the amount of SG stars
formed, but also their overall chemical distribution, as the chemical
abundances of the FG ejecta vary in time (cf. Fig. \ref{fig:tab}). Its
value is constrained by the observational evidence (discussed in
sect. \ref{subsec:cno}) that the sum of CNO elements varies no more
than a factor two among the GC stars \citep{coh05} \citep[or by a factor
three in NGC~1851, see][] {ventura18512009}; this implies that the SG
progenitors must be AGB stars with masses larger than $\sim$5 $\msun$
($\sim 4$ $\msun$ in the latter case) and that the SG formation can not
continue longer than the lifetime of these stars, that is $t_{\rm
  end}\la 100$ Myr (or $\la 150$\ in the latter case) \footnote{A
  possible mechanism able to clear the ISM in a globular clusters in
  this lapse of time is given by the explosions of type Ia supernovae
  (see Paper I).}  (cf. Fig. \ref{fig:tab} for the behaviour of C+N+O
with time).

Finally, we introduce the ratio $x=\rho^{\rm n}_{\rm *,SG}/\rho_{\rm
  *,tot}$ between the SG stars and the total nowadays alive stars
$\rho_{\rm *,tot}=\rho^{\rm n}_{\rm *,FG}+\rho^{\rm n}_{\rm *,SG}$.
Its current value depends not only on the duration and efficiency of
the SF of the SG stars, but also on the subsequent loss of FG stars
(see Paper I). We point out that this parameter does not enter
directly into the models (it is in fact absent in
eqs. \ref{eq:den}-\ref{eq:str}); its value is inferred {\it a
  posteriori} for a more realistic fit of the data.  We will return on
this point in sect. \ref{sec:comparison}, when comparing our models to
specific GCs.

In the end, we summarize in Table \ref{param} the key parameters
(besides the chemistry of the AGB ejecta) characterizing our models.

\begin{table}
\caption{Parameters characterizing the models}             
\label{param}      
\begin{tabular}{ l l}     
\hline
$t_{\rm ac}$  & time of maximum accretion of  pristine gas  \\
$\tau$        & time scale of the  pristine gas accretion  \\
$t_{\rm end}$ & time marking the end of the SG formation phase \\
$\rho_{\rm *,FG}$ & density of the FG stars \\ 
$\rho_{\rm 0,pr}$ & density regulating the amount of available pristine gas \\
$\nu$        & SF efficiency \\
$x$          & the ratio between the nowadays alive SG and total stars \\
\hline 
\end{tabular}                                                              
\end{table}

\section{Model analysis}
\label{sec:moana}

\begin{figure}    
\centering{
\includegraphics[width=8cm]{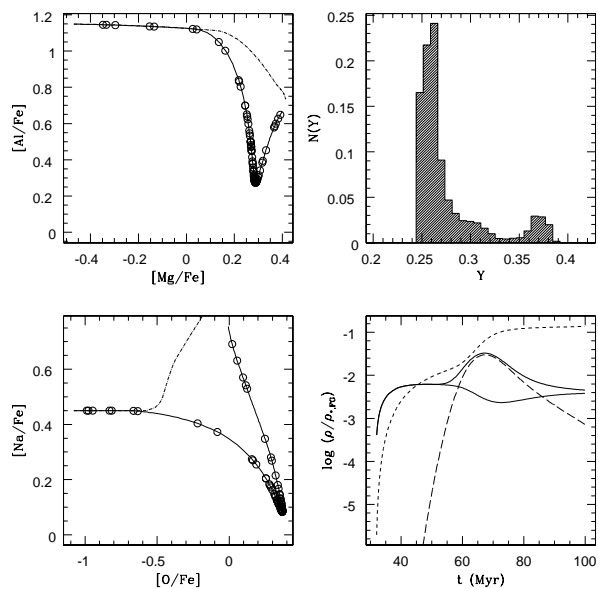}
}
\caption{Top-left panel: chemical evolution of the RM in the plane
  [Mg/Fe]-[Al/Fe]. Bottom-left panel: chemical evolution of the RM in
  the plane [O/Fe]-[Na/Fe] plane. In these panels the
  dot-dashed line represents the ISM chemical pattern of the model
  without accretion of pristine gas (RMNOPG), while the solid line
  holds for the RM, in which the accretion occurs; the open circles
  represent a statistical sampling of the SG stars.  Top-right panel:
  helium distribution function of the SG stars.  Bottom-right panel:
  evolution of the total amount of ISM (bold solid line), AGB ejecta
  (solid line), pristine gas (long-dashed line) and SG stars
  (short-dashed line).  }
\label{fig:refmod} 
\end{figure}

\subsection{The reference model}
\label{subsec:rm}

To illustrate how our model works, we focus our attention on a model
with the following parameter set $(t_{\rm ac,7}, t_{\rm
  end,7},\tau_7,\rho_{\rm 0,pr},\rho_{\rm
  *,FG},\nu,x)=(6.5,10,0.6,15,150,0.5,1)$, where the times are
expressed in $10^7$ yr and the densities in $\msun$ pc$^{-3}$.  Note
that we here are interested only in the evolution of the SG stars, and
we set $x=1$. We checked that the current choice of parameters gives
(in absence of the accretion of pristine gas) the same temporal
profile of the gas and of the SG stars of the reference model of Paper
I (cf. its Fig. 2); for this reason, we'll refer to the one zone model
described in this section as the reference model (RM). We consider an
amount of available pristine gas one tenth of the FG mass, nearly a
factor two larger than that of the ejecta delivered after $\sim 10^8$
yr.
 
The main properties of this model are shown in
Fig. \ref{fig:refmod}. The two left panels illustrate the chemical
path of the model in the planes [Al/Fe]-[Mg/Fe] and
[O/Fe]-[Na/Fe]. The top-right panel provides the helium distribution
function (hereafter HDF) of the SG stars, while the bottom-right panel
shows the time evolution of the amount of several quantities.
Initially, only gas returned by the FG stars is present (thin solid
line). Later on, the total gas content (thick) first experiences a
``jump'' as a consequence of the pristine gas (long-dashed line)
accretion, and successively decreases because of the star
formation. The amount of SG stars (short-dashed line) shows, as
expected, a rapid increase in conjunction with the accretion.

\begin{figure}    
\centering{
\includegraphics[width=8cm]{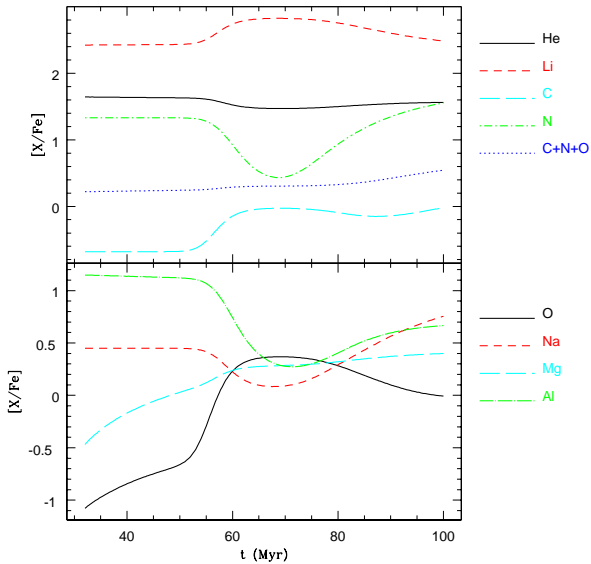}
}
\caption{Time evolution of the chemical abundances in the ISM
of the RM.}
\label{fig:refchm} 
\end{figure}

Figure \ref{fig:refchm} illustrates the abundance evolution of different
elements in the ISM of the RM. A comparison of Fig.  \ref{fig:tab}
and Fig. \ref{fig:refchm} clearly shows that, as anticipated in
sect. \ref{subsec:ona}, the presence of pristine gas is necessary in
order to reproduce the O-Na anticorrelation observed in all the GCs.
In Fig. \ref{fig:tab} the O and Na abundances in the FG stellar
ejecta grow simultaneously, and no anticorrelation between these two
elements can be established; on the other hand, Fig. \ref{fig:refchm}
clearly indicates how the pristine gas accretion produces opposite
slopes in the time evolution of the O and Na abundances in the ISM,
thus producing an anticorrelation similar to that observed in globular
clusters.

This point is illustrated even more vividly in the bottom-left panel
of Fig. \ref{fig:refmod}. This panel includes the chemical path of the
gas {\it in absence of any accretion} (i.e. $\rho_{\rm pr}=0$),
represented by the dot-dashed line. This model (hereafter the RMNOPG
model) starts from the O-poor end of this line since the ejecta of the
most massive AGBs are assumed to be particularly deprived of oxygen
(see Fig. \ref{fig:tab} and the discussion in
sect. \ref{sec:yelds}). As gas from less massive AGB stars contributes
to the ISM both the abundances of Na and O increase giving rise to a
correlation rather than the observed anticorrelation.  The path
followed by the RM when the effect of pristine gas accretion is
included clearly illustrates the role of this gas in establishing the
Na-O anticorrelation. Since the ISM is initially composed only by AGB
ejecta, the RM track initially coincides with that of the model with
no pristine gas. At $t\sim 55$ Myr, when [O/Fe]$\ga -0.5$, the amount
of the infalling pristine gas becomes substantial, and the RM moves
``downward'' at increasing values of [O/Fe], giving rise to a
[O/Fe]-[Na/Fe] anticorrelation. This motion is not uniform because the
pristine gas accretion suddenly increases (decreases) the abundance of
oxygen (sodium), ``pushing'' the model in short time towards higher
(lower) values of [O/Fe] ([Na/Fe]) (cf. Fig.  \ref{fig:refchm}).  As
the amount of pristine gas in the ISM increases, the trajectory moves
toward its minimum located at [O/Fe]=0.37 and [Na/Fe]=0.1 dex,
``above'' the point identified by the ratios [O/Fe]=0.4 and [Na/Fe]=0
of the pristine gas.  Successively, the accretion rate of the pristine
gas reduces becoming lower than the SFR, and the relative amount of
this gas in the ISM decreases more and more. The ISM composition
evolves back toward the chemical abundance of the AGB ejecta and the
chemical path of this model tends again toward that of the RMNOPG
model.

 It is important to point out that the ISM trajectory in the Na-O plane
is not sufficient to predict the distribution of SG stars in this
plane. In fact, such a path is, in a sense, similar to a stellar
Color-Magnitude diagram in which some regions are more populated than
others owing to the longer time spent there by the stars. In our model
the amount of SG stars along a certain portion of the chemical
trajectory depends on the rate of evolution of the ISM chemical
abundance as well as on the SFR (which is proportional to
$\rho^{3/2}$).  In order to show how the [O/Fe]-[Na/Fe] and
[Mg/Fe]-[Al/Fe] diagrams are actually populated by the SG stars, in
Fig. \ref{fig:refmod} we have plotted twenty open dots representing a
statistical sampling of the SG stars forming along the chemical path.
Consider, for instance, the diagram [O/Fe]-[Na/Fe]: only a few stars
populate the two branches of the curve, while most of them form close
to its cusp; the cusp coincides with the maximum amount of ISM and the
consequent maximum SFR. Although the time spent by the model along
the two branches is similar (cf. Fig. \ref{fig:refchm} and the
bottom-right panel of Fig. \ref{fig:refmod}), the descending one is
more populated by stars because the ISM density and the SFR are larger
at this stage.

Finally, the top-right panel of Fig. \ref{fig:refmod} shows the HDF. A
small number of stars (10\%) have an helium abundance $Y>0.35$. These
stars coincide with the most O-poor stars present in the
[O/Fe]-[Na/Fe] diagram. The stars giving rise to the peak around the
range $0.24<Y<0.29$ form when the maximum amount of pristine gas is
present in the GC, and correspond to the stars populating the region
 around the cusp. The tail at $Y\sim 0.3$ is composed by stars formed
during two nearly chemically ``symmetric'' phases, just before and
after the occurrence of the bulk of the pristine gas accretion (cf. Fig.
\ref{fig:refchm})).

\subsection{Exploring the parameter space}
\label{subsec:space}
We now explore the dependence of the model on the chosen set of
parameters through variations of the chemical paths in the
[O/Fe]-[Na/Fe] plane and variations of the HDF. We
consider several models, each differing from the RM in the value
of a single parameters. All the models are listed in Table \ref{paramodel}.

\begin{table*}
\caption{Models adopted for the exploration of the parameter space.}             
\label{paramodel}      
\centering          
\begin{tabular}{ l c c c c c c r }     
\hline
  Models & $t_{\rm ac,7}$ & $\tau_7$ & $t_{\rm end,7}$ & $\rho_{\rm
    pr}$ & $\rho_{\rm *,FG}$ & $\nu$ & panel and style$^c$\\
\hline
 RM      & 6.5           & 0.6     &  10.0          & 15   & 150  & 0.5 &  all, solid, black\\
 RMNOPG  &               &         &                & 0.0         & & & all, dot-dashed, green\\
 RMMPRL  &               &         &                & 4.5          & & &$a$, dashed, blue   \\
 RMMPRM  &               &         &                & 45        & & & $a$, dotted, red        \\
 RMTACL  & 5.0           &         &                &              & &  & $b$, dashed, blue       \\
 RMTACM  & 8.0           &         &                &              & &  & $b$, dotted, red       \\
 RMTAUL  &               & 0.3     &                &              &   &   & $c$, dashed, blue    \\
 RMTAUM  &               & 1.2     &                &              &   &  & $c$, dotted, red      \\
 RMEFFL  &               &         &                &            & &  0.1   & $d$, dashed, blue   \\
 RMEFFM  &               &         &                &           &  &  1.0   & $d$, dotted, red   \\
 RMTNDL  &               &         &  7.0           &         &    &      & $e$, dashed, blue     \\
 RMTNDM  &               &         & 13.0           &        &     & & $e$, dotted, red      \\
 RMFGDL   &               &         &                  &   1.5   & 15 & &$f$, dashed blue   \\
 RMFGDM   &               &         &                  &    150 & 1500  & &$f$, dotted, red    \\
 RMTSFL  &               &         &                &        &  &  0.0 for $t_7<5$ & dashed, blue$^d$   \\
         &               &         &                &        & & 1.0 for $t_7\gid 5$ & \\
 RMTSFM  &               &         &                &       &    &  0.0 for $t_7<7$ & dotted, red$^d$   \\
         &               &         &                &      &  &  1.0 for $t_7\gid 7$&\\

\hline 
\end{tabular}  
\par\noindent  
                       
\leftline{$^a$Times are given in $10^7$ yr and densities in $M_{\odot}$ pc$^{-3}$.}
\leftline{$^b$For each model only the parameter values differing from the RM are reported, all
  the others being the same.}
\leftline{$^c$This column indicates the panels of Figs. \ref{fig:mix} and
  \ref{fig:imix} in which the single models are shown, and their line
  style.}
\leftline{$^d$Referred to Fig. \ref{fig:delsf}.}
\end{table*}

\begin{figure*}
\centering{
\includegraphics[width=18cm]{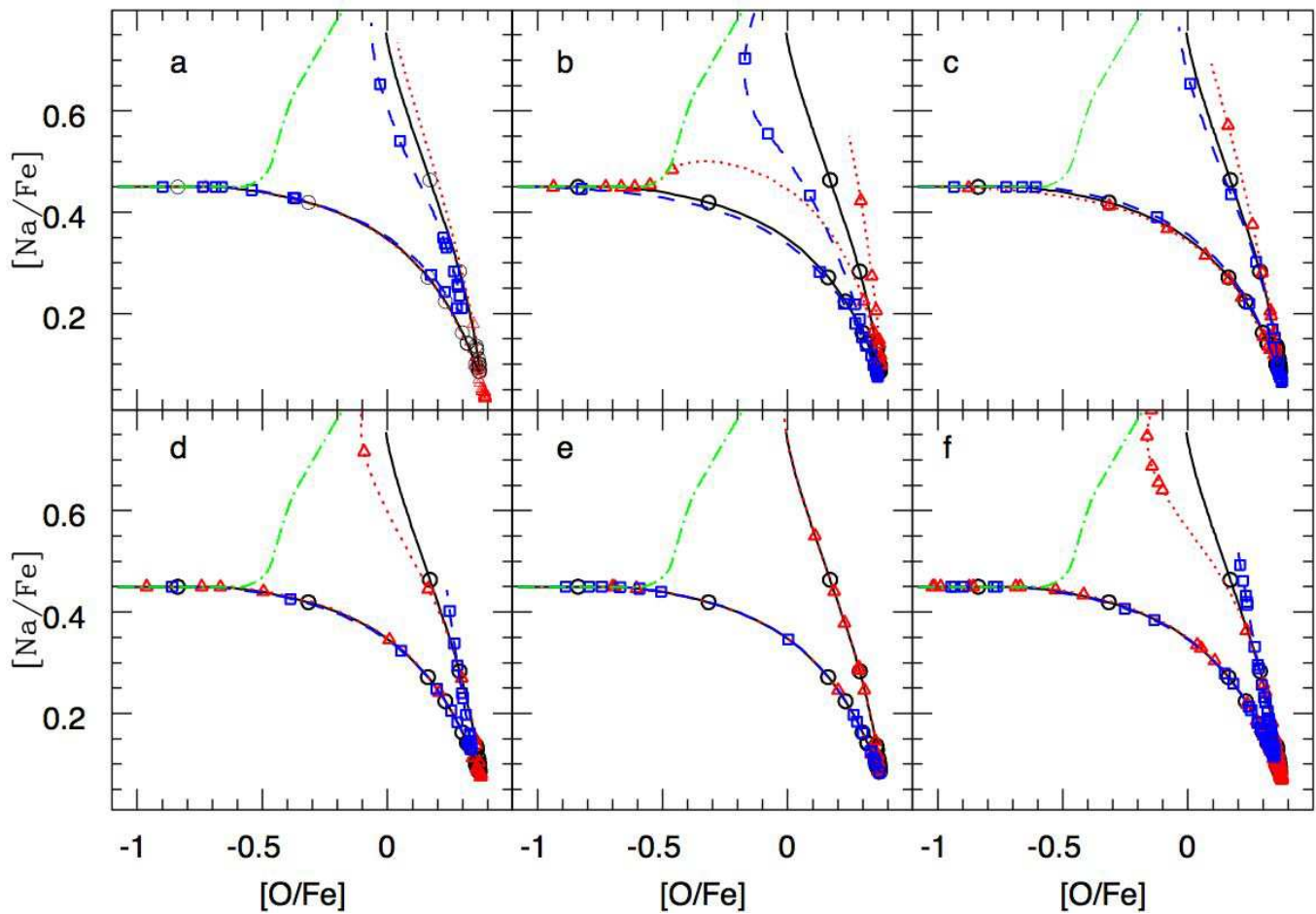}
}
\caption{Chemical paths of models with modified parameters with
  respect to the RM. Each panel highlights the effect of a single
  parameter (see the text for more details). For practicality's sake,
  the RM (black solid line) and RMNOPG (green dot-dashed line) are
  reported in all panels. The different symbols represent statistical
  samples of the SG stars in different models.  }
\label{fig:mix} 
\end{figure*}

\subsubsection{The [O/Fe]-[Na/Fe] diagram}
\label{subsubsec:onadiag} 
The six panels in Fig. \ref{fig:mix} illustrate how changes of each
single parameter affect the RM behaviour in the [O/Fe]-[Na/Fe]
diagram. The RM path (solid line) is replicated in every panel to
facilitate the comparison with the different modified models. For the
same reason we also include in all the panels the dot-dashed line
representing the behaviours of the RMNOPG (see Fig. \ref{fig:refmod}).
We now discuss these panels.
 
\begin{description}
\item[{\it Panel a} -] In this panel the RM is compared with two
  similar models, but with $\rho_{\rm pr}=4.5$ $\msun$ pc$^{-3}$
  (RMMPRL, blue dashed line) and $\rho_{\rm pr}=45$ $\msun$ pc$^{-3}$
  (RMMPRM, red dotted line). The O-poor branch of the RMMPRL
  substantially overlaps the analogous RM branch as long as the amount
  of accreted pristine gas is relatively low in both models. The
  rising branch, instead, is shifted toward lower values of
  [O/Fe]. This shift is a consequence of the smaller amount of
  pristine gas, and its lower influence on the chemical abundance of
  the ISM; the dashed line tends to return more rapidly toward the
  dot-dashed one. The reduced value of $\rho_{\rm pr}$ affects the
  position of the cusp, which occurs at [O/Fe]=0.3 and [Na/Fe]=0.2
  earlier than in the RM model.

  Also in this case the stars (open blue squares) form mainly around
  the cusp, but the two branches are more populated. This is due to
  the combined effect of the lower amplitude of the gaussian accretion
  rate and the dependence of the SFR on $\rho^{3/2}$; the result is a
  less marked peak in the SFR occurring at the cusp.

  Similar considerations apply to the RMMPRM.
  While the descending branch overlaps again with those of the previous
  models, the cusp of the red dotted curve is located at a value of [Na/Fe]
  nearly 0.2 dex lower than that of the blue curve, and the rising
  branch is shifted toward higher values of [O/Fe]. Finally, in this
  model the stars (open red triangles) are essentially all
   concentrated around the cusp.

\item[{\it Panel b} -] The models represented here are similar to the
  RM, but with $t_{\rm ac,7}=5$ (RMTACL, blue dashed line) and $t_{\rm
    ac,7}=8$ (RMTACM, red dotted line).  The O-poor descending branch of the
  dashed line is essentially similar to that of the RM, but slightly
  shifted toward lower values of [Na/Fe]; this is a consequence of the
  earlier accretion of pristine gas. The rising branch is instead
  longer and strongly shifted toward lower values of [O/Fe] because at
  this stage the amount of pristine gas eroded by the SF is larger
  than in the RM. The dashed line tends therefore to reconnect earlier
  to the RMNOPG chemical pattern.

  Given the delay in the gas accretion, the RMTACM follows the
  RMNOPG track for a longer interval, then quickly converges toward
  the cusp. The dotted rising branch is shorter and shifted toward
  higher values of [O/Fe] for opposite reasons relative to the RMTACL.

  Because of the delay in the accretion, the RMTACM descending branch
  is more populated by stars than the rising branch, while the
  opposite happens for the RMTACL. Once more, this is due to the
  fact that the SFR grows with the gas density. Note the gap in the
  stellar distribution along the descending dotted branch (open
  triangles). Initially, the RMTACM has enough time to form a
  substantial amount of AGB ejecta before the accretion becomes
  significant. A non negligible fraction of O-poor stars can thus
   form. The subsequent accretion moves the model toward the cusp very
  rapidly, and very few stars can be produced during this stage.

\item[{\it Panel c} -] In this panel the models with $\tau_7=0.3$
  (RMTAUL, blue dashed line) and $\tau_7=1.2$ (RMTAUM, red dotted
  line) are compared. Smaller values of $\tau$ correspond to a more
  rapid accretion. The two lines are rather similar, but with the
  dotted rising arm being shorter and shifted toward O rich values; in
  this case, in fact, the longer duration of the accretion produces a
  more persistent influence on the RMTAUM which thus shows some delay
  in its tendency to reconnect with the RMNOPG line.

  The distributions of the stars along the dashed and dotted lines are
  also similar. However, a prevalence of O-poor stars in the
  RMTAUL (open squares) is present; owing to the shorter extent
  of the accretion, a larger amount of AGB ejecta can be initially
  collected without being significantly diluted by the pristine gas.
 
\item[{\it Panel d} -] The effects of the SF efficiency are
  investigated by models with $\nu=0.1$ (RMEFFL, blue dashed line) and
  $\nu=1$ (RMEFFM, red dotted line). As usual, the descending branches
  essentially overlap. The rising RMEFFM arm, instead, is more
  extended and shifted toward lower O abundances because the larger SF
  efficiency leads to a more rapid consumption of the ISM in general,
  and of the pristine gas in particular; once the bulk of the
  accretion is finished, the fresh AGB ejecta is diluted by a lower
  amount of the pristine gas still present in the ISM, and the RMEFFM
  chemical pattern moves more quickly toward the RMNOPG track.

 \item[{\it Panel e} -] In all the previous models, we assumed $t_{\rm
     end,7}=10$. Obviously, the duration of the SG formation phase may
   be different, and we studied the influence of such a duration on
   the chemical path of the SG stars. In this panel we show a model
   with $t_{\rm end,7}=7$ (RMTNDL, blue dashed line) and a model with
   $t_{\rm end,7}=13$ (RMTNDM, red dotted line). As expected, the
   chemical paths coincide, but with the dotted one obviously longer
   than the dashed one. As the two models are identical, the number of
   stars formed along the common tracks is the same, and one could
   expect the same distribution of triangles and squares on
   them. Instead, it is apparent that the triangles are less than the
   squares on the first branches; this is due to the fact that the
   number of symbols indicating the stars is not proportional to the
   total number of stars actually formed; rather, for every model we
   adopt twenty symbols to represent the total number of SG stars in
   it, no matter if this number is different in different models. In
   particular, the two models shown in this panel are identical, and
   the same number of stars is formed along the descending branches;
   however, the RMTNDM lasts longer, and many more stars are formed
   along the rising branch.  For this reason more triangles are
   distributed on this branch and less on the descending one, where
   the squares are more numerous.
 
 \item[{\it Panel f} -] In this panel we show the behaviour of two
   models having both the values of $\rho_{\rm *,FG}$ and $\rho_{\rm
     pr}$ ten times lower (RMFGDL, blue dashed line) and ten times
   higher (RMFGDM, red dotted line) than in RM.  In
   sect. \ref{subsec:rm} we have justified our choice of $\rho_{\rm
     *,FG}=150$ $\msun$ pc$^{-3}$; this choice, however, although
   reasonable, remains arbitrary. We expect that our results depend on
   a scaling of $\rho_{\rm *,FG}$ and $\rho_{\rm pr}$, even if their
   ratio, and thus the relative amount of dilution due to the pristine
   gas, remains the same; in fact, a lower (higher) density of
   $\rho_{\rm *,FG}$ produces a lower (higher) density of the AGB
   ejecta which, together with a lower (higher) $\rho_{\rm pr}$,
   determine a lower (higher) SFR and a different evolution.

   This difference is apparent in the present panel.  As usual, also
   the two models shown here have very similar descending branches,
   but the RMFGDL cusp occurs at a higher value of [Na/Fe]. This is
   because, owing to the lower SFR in this model, less ejecta are
   initially subtracted by the star formation, and the dilution by the
   pristine gas is less effective.  Successively, however, always
   because of the lower SFR, the pristine gas is depleted more slowly,
   remaining longer in the ISM and leading to a shorter rising
   branch. Opposite arguments hold for the RMFGDM, whose cusp is
   ``deeper'' and whose rising branch is longer. Note also that this
   arm is strongly bent ``to the left'' because, given the larger
   amount of ejecta (relative to the pristine gas) in this case, this
   model tends more quickly toward the RMNOPG.
\end{description} 

\begin{figure*}
\centering{
\includegraphics[width=18cm]{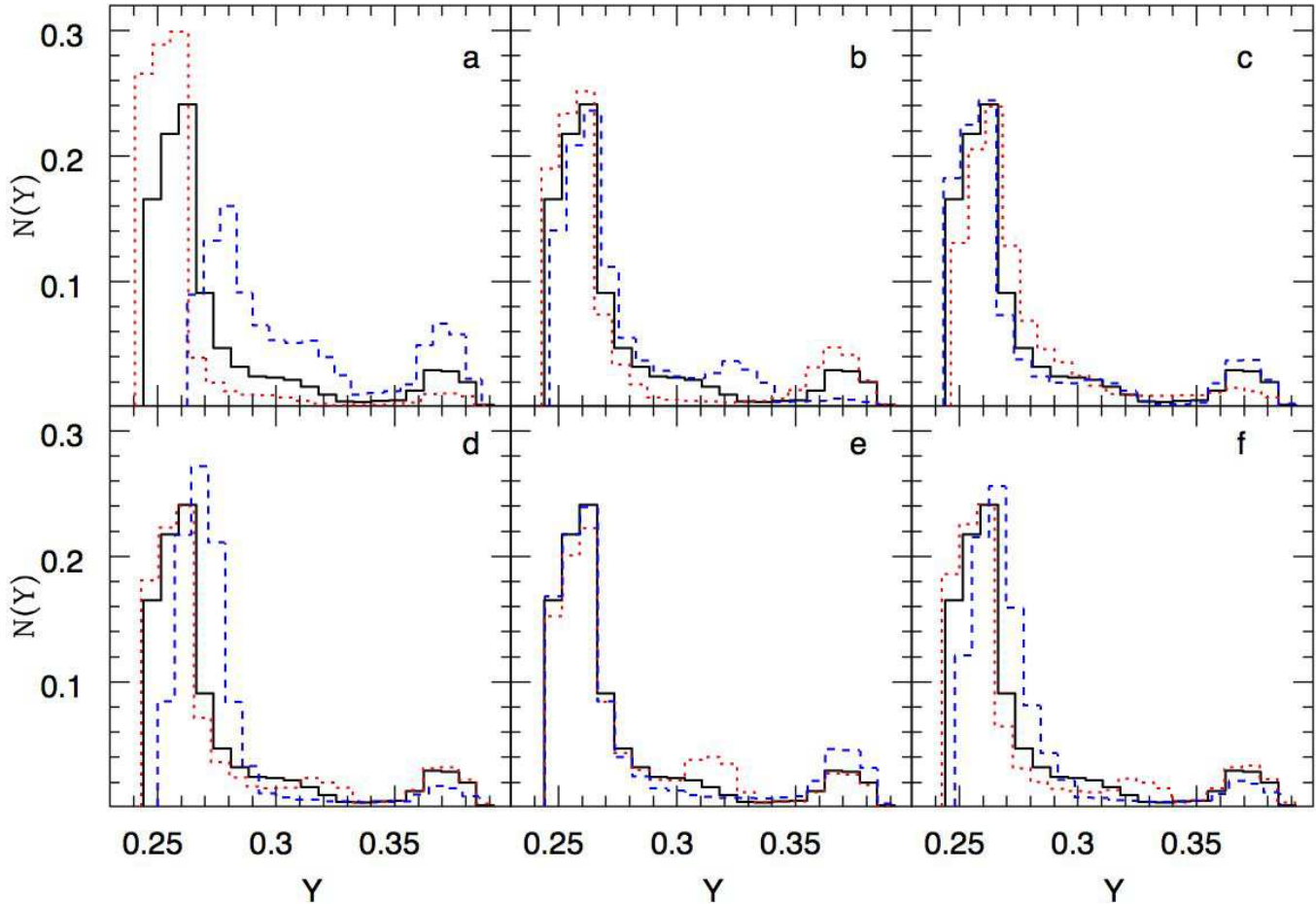}
}
\caption{Stellar helium distribution function for different
   models. The models in each panel are the same shown in the
  corresponding panel of Fig. \ref{fig:mix} }
\label{fig:imix} 
\end{figure*}

\subsubsection{The helium distribution function}
\label{subsubsec:hdf} 
Another crucial output of the models are the stellar HDFs, as some
observed globular clusters show different populations separated by
sharp differences in helium abundances.  The HDFs of the models
illustrated in Fig. \ref{fig:mix} are shown in Fig. \ref{fig:imix}. We
now briefly discuss this figure:

\begin{description}
\item[{\it Panel a} -] This panel shows that in the RMMPRM the vast
  majority of the stars have $Y \sim 0.25$, with a thin tail extending
  up to $Y=0.37$. For smaller amounts of pristine gas (dashed
  histogram) the left peak is lowered and shifted toward larger values
  of $Y$ (for the same reason, the cusp of the dashed line in panel
  $a$ of Fig. \ref{fig:mix} is shifted compared to the RMMPRM cusp).
  The tail in the range $0.29<Y<0.33$, as well as the
  secondary peak at $Y=0.37$, increase their relative importance.

\item[{\it Panel b} -] The temporal shift of pristine gas accretion
  produces its major effects on the secondary, helium rich peak. If
  the accretion is delayed (RMTACM, dotted line), a larger number of stars
   form with the pure, He-rich, AGB ejecta, giving rise to a
  non-negligible secondary peak at $Y=0.37$. With an early accretion
  (dashed line) this peak is substantially erased because only few
  stars form with such an high $Y$ value as the system quickly
  accretes pristine gas forming the main peak at $Y\sim 0.26$. Later
  on, as discussed in the previous subsection, most of the pristine
  gas is depleted by the SF and the ISM is composed of a
  larger fraction of He-rich AGB ejecta; thus the stars forming at
  this stage give rise to the HDF secondary peak at $Y\sim 0.32$.

\item[{\it Panel c} -] From panel $c$ of Fig. \ref{fig:mix} we have
  seen that in the model with larger $\tau$ (RMTAUM, dotted line) the
  cusp is slightly brought forward; for the same reason the main peak
  of the RMTAUM HDF is shifted a little ``rightward''. The greater
  height of the secondary peak at $Y=0.37$ in the RMTAUL (dashed
  histogram) is due to the longer time during which the stars form
  from pure AGB ejecta, before the accretion becomes substantial.

\item[{\it Panel d} -] The RMEFFL 
  (dashed histogram) creates little less O-poor, He-rich stars at early
  times, and the HDF shows therefore a slightly lower secondary peak
  at $Y=0.37$ A smaller $\nu$ leads to a longer persistence of the
  pristine gas and of its effect on the ISM chemical
  characteristics. For this reason the cusp of the dashed curve is
  shifted ``upward'' in panel $d$ of Fig. \ref{fig:mix}, and the
   dashed HDF is shifted ``rightward'' in the present panel.

\item[{\it Panel e} -] The models shown in panel $e$ differ by the
  evolutionary time. Although the chemical paths are identical, the
  HDFs are quite different and reflect the different stellar crowding
  of the two arms (cf. panel $e$ of Fig. \ref{fig:mix}). The RMTNDL
  histogram (shorter evolution) shows two well defined populations: a
  minor one with $Y>0.35$, and the major one peaked around
  $Y=0.26$. In the RMTNDM histogram a third population is present at
  $Y=0.31$ as a consequence of the longer permanence of the model on
  the rising branch.

\item[{\it Panel f} -] Given the larger amount of AGB ejecta in the
  RMFGDM (red dotted line), its He-rich peak is higher than in the
  RMFGDL. The dotted histogram also shows an intermediate population in
  the range $0.3<Y<0.35$ which is substantially absent in the dashed
  histogram. This intermediate population is a consequence of the
  bending toward lower values of O of the rising branch of the RMFGDM
  in the [O/Fe]-[Na/Fe] plane.
\end{description}

\subsubsection{Delayed star formation}
\label{subsubsec:delsf} 
As pointed out in \ref{subsec:supagb}, in some clusters such as $\omega$
Cen and NGC 2808 multiple discrete populations are present, each
characterized by a specific helium abundance. This led
\citet{ren08} to argue that helium-rich material was
accumulated in the ISM for a sufficient long time until suddenly a
burst turned a major fraction of the ISM into stars.
According to this author, a continuous SF proceeding along with the
ISM helium enrichment would have resulted in a continuous distribution
of helium abundances in the newly formed stars, hence in a broadening
of the MS rather than in well separated sequences as actually
observed.

\begin{figure}    
\centering{
\includegraphics[width=8cm]{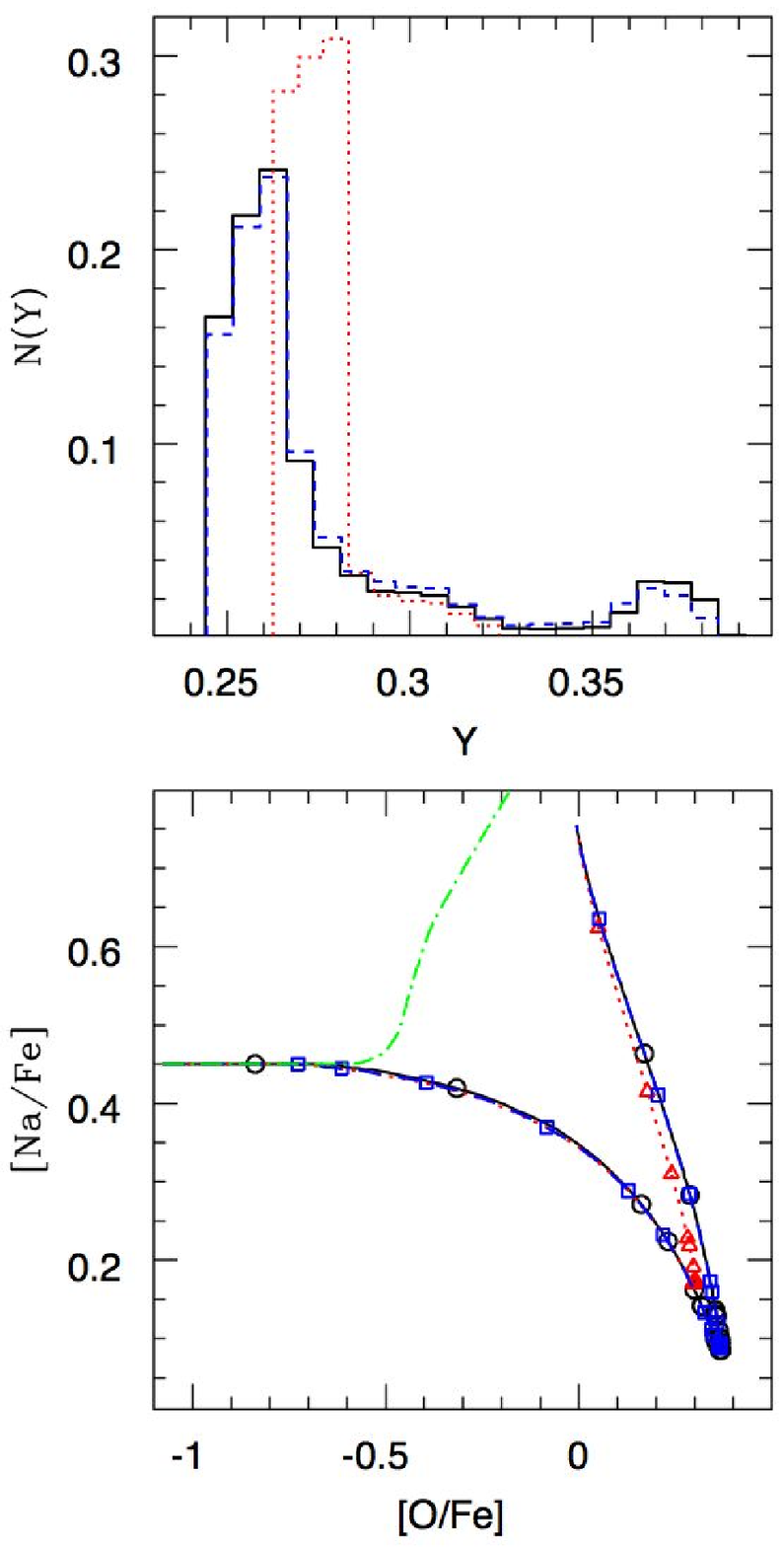}
}
\caption{Helium distribution function (upper panel) and chemical
  pattern (lower panel) of RM (solid line), RMTSFL (blue dashed line)
  and RMTSFM (red dotted line). The green dot-dashed line indicates
  the evolution of RMNOPG.}
\label{fig:delsf} 
\end{figure}

In all the models we discussed so far the SF starts ``ab initio'', as
soon as the ejecta of the massive AGB stars begins to collect within
the GC. To investigate the effect of a SF burst we run two models in
which the gas cumulates in the GC since the beginning as usual, but
with the SF inhibited up to the time $t_{\rm b,7}=5$ (RMTSFL, blue
dashed line) and $t_{\rm b,7}=7$ (RMTSFM, red dotted line). The
chemical patterns are very similar (see the lower panel in
Fig. \ref{fig:delsf}), but with the dotted cusp shifted 0.1 dex
``leftward'' and ``upward'' compared to the dashed one. In fact, in
the RMTSFM a large amount of AGB ejecta gathers before the occurrence
of the SF; the dilution by the pristine gas is therefore less
effective and the cusp takes place earlier. In this latter model the
stars are essentially distributed along the rising branch, as expected
given the large SF delay.

In RMTSFM the effects of the pristine gas are less effective, and the
primary peak of the HDF (see the upper panel in Fig. \ref{fig:delsf})
is higher and shifted rightward (reflecting the shift of the cusp of
dotted line in the lower panel of Fig. \ref{fig:delsf}). The secondary
peak is absent because the SF was not active in the beginning, when
the gas was very helium-rich.  This simple experiment shows that, with
a SF delay large enough, it is possible to obtain a sharp intermediate
population, but to the expense of the extreme one. Both populations
can not be achieved with this mechanism.

\section{Comparison with real clusters}
\label{sec:comparison}

In this section we apply the chemical evolution framework described in
the previous section to build two specific models for NGC 2808 and
M4. These two clusters are indeed good prototypes to exemplify the
cluster to  cluster differences. NGC~2808 is a complex mixture of
 populations, one of which  is very extreme, and has a very high helium
content testified by the presence  of the blue main sequence. M4 is
much less massive, and shows milder variations  in elemental
abundances, particularly in oxygen, but it apparently has a larger
spread in sodium. It is to be preferred to other samples also because
a very  accurate recent spectroscopic survey \citep{mar08}

\subsection{NGC2808} 
\label{sec:ngc2808}

The cluster NGC 2808 has a mass $M\sim 1.6\times 10^6$ $\msun$, and
harbours three main populations, as initially speculated by
\citet{daca04} and successively confirmed by \citet{pio07}. Each
population has a distinct helium abundance: $\sim 50$\% of the stars
are credited with a primordial helium abundance ($Y=0.25$), $\sim
32$\% with $Y\sim 0.30$ and $\sim 18$\% with $Y\sim 0.37$
\citep{carretta2009a,dc2008}.  No iron abundance difference is present
in these populations, but they show an extended [O/Fe]-[Na/Fe]
anticorrelation \citep{car06}.

\begin{figure}    
\centering{
\includegraphics[width=8cm]{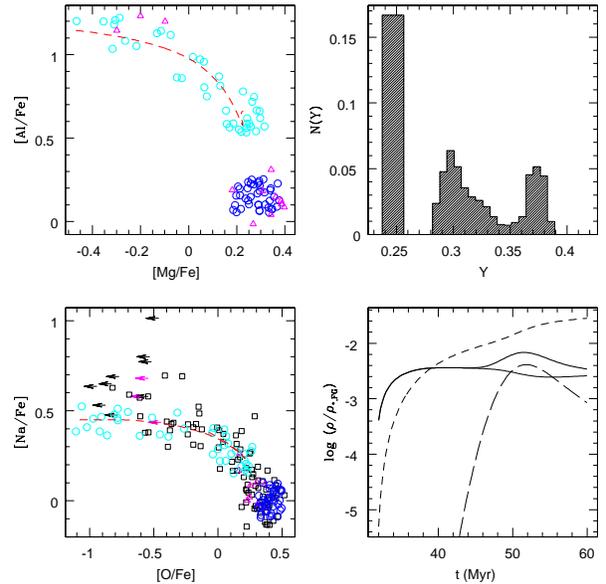}
}
\caption{Results of our model for NGC2808. The following set of
  parameters has been assumed: $(t_{\rm ac,7},t_{\rm
    end,7},\tau_7,\rho_{\rm pr},\rho_{\rm
    *,FG},\nu,x)=(5,6,0.3,2.36,241,1,0.5)$.  The top-left panel and
  the bottom-left panel show the [Mg/Fe]-[Al/Fe] and the
  [O/Fe]-[Na/Fe] diagrams relative to NGC 2808, respectively. The
  black squares and arrows are data obtained with the ESO
  high-resolution multifiber spectrograph FLAMES/GIRAFFE at VLT by
  \citet{car06}. Magenta triangles and arrows are data obtained with
  the higher resolution UVES spectrograph \citep{carretta2009b}. The
  blue and cyan circles represent a sampling of the FG and SG stars by
  our model, respectively. The red dashed line represents the gas
  trajectory within the diagram; the sampled stars in principle would
  be located on this line, but we introduced a random scatter in the
  range 0 - 0.1 dex in their coordinates in order to mimic the
  observational errors (see text for more details). The top-right
  panel illustrates the stellar helium distribution function. In the
  bottom-right panel the evolution of the following quantities is
  reported: total amount of gas (bold solid line), stellar ejecta
  (solid line), pristine gas (long-dashed line), SG stars (dashed
  line).  }
\label{fig:2808} 
\end{figure}

This anticorrelation is apparent in the bottom-left panel of
Fig. \ref{fig:2808} which includes data obtained with the ESO
high-resolution multifiber spectrograph FLAMES/GIRAFFE at VLT by
\citet{carretta2009a} and data obtained with the higher resolution UVES
spectrograph \citep{carretta2009b}.

\begin{figure}    
\centering{
 \includegraphics[width=8cm]{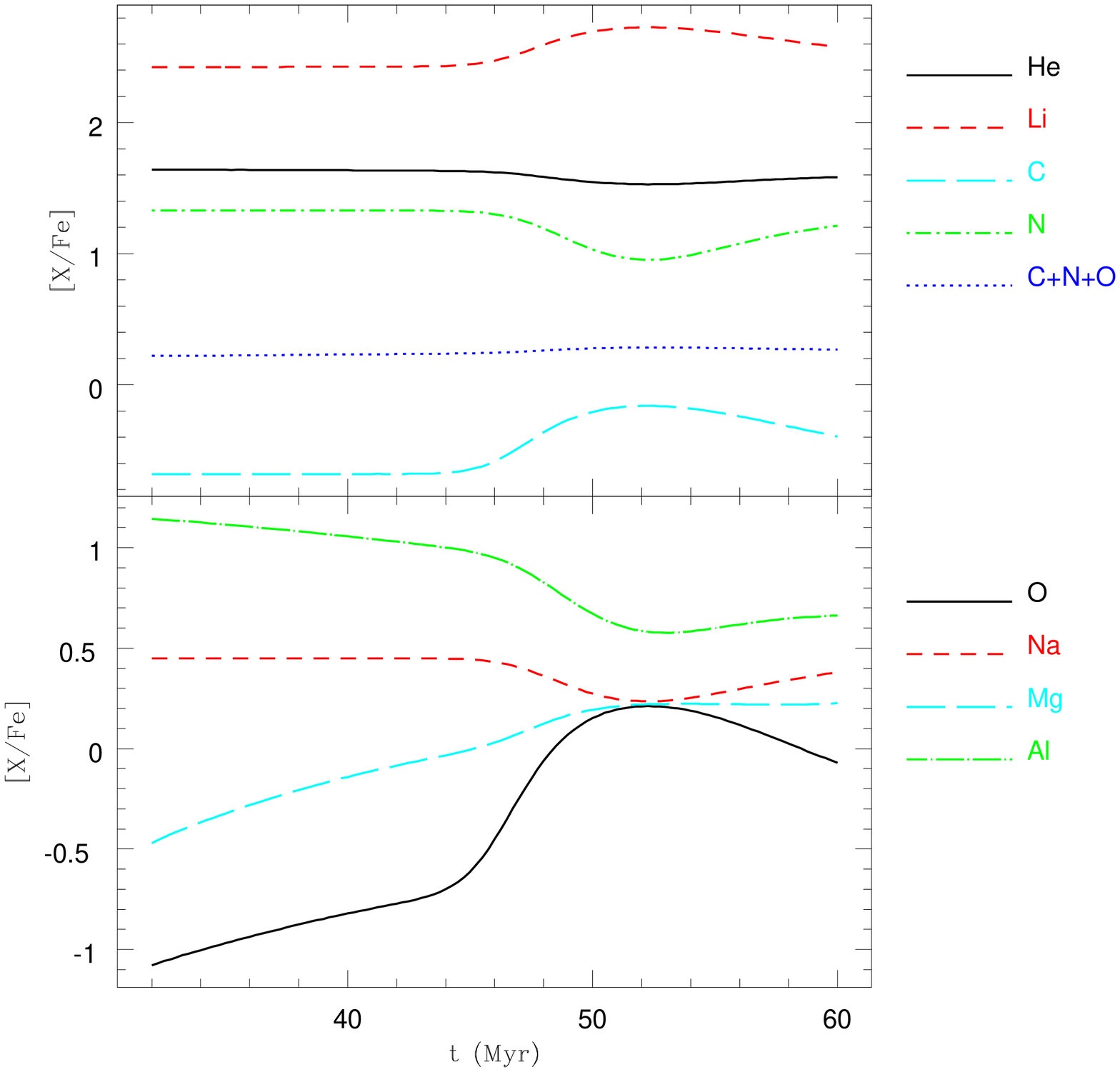}
}
\caption{Time evolution of the chemical abundances in the ISM of 
the NGC 2808 model.
}
\label{fig:chm2808} 
\end{figure}

We simulated the GC NGC 2808 with a model with the parameter set $(
t_{\rm ac,7},t_{\rm end,7},\tau_7,\rho_{\rm pr},\rho_{\rm
  *,FG},\nu,x)=(5,6,0.3,2.36,241,1,0.5)$. The choice of $x=0.5$ is
done in order to get a current 50\% of FG stars, as suggested by the
helium abundance; this produces a more realistic comparison with the
observations.

The time evolution of the ISM chemical composition is illustrated in
Fig. \ref{fig:chm2808}; as shown in this figure, owing to the relative
shortness of the evolutionary time the C+N+O abundance remains
essentially constant in this model. We remind that the absence of a
substantial CNO increment is an essential constrain for any model
describing the chemistry of the majority of GCs
(cf. sect. \ref{subsec:cno}).

The cyan circles in the bottom-left panel of Fig. \ref{fig:2808}
represent a statistical sample of the SG stars obtained integrating
the equation system in sect. \ref{sec:comp}. In populating the
[O/Fe]-[Na/Fe] plane with such circles, we introduce a stochastic
error in the range 0-0.1 dex, comparable to the observational errors,
scattering the position of the stars.

The blue circles represent the FG stars. Since the FG stellar
population is assumed to be present ``ab initio" in our model, these
stars are not an outcome of the model, but are included in the figure
to ease the comparison with the observational data. The pristine gas
from which these stars formed is assumed to have [O/Fe]=0.4 and
[Na/Fe]=0 and also in this case we introduce a stochastic error in the
range 0-0.1 dex in the abundances. The relative number of FG and SG
stars is given by the value of the parameter $x$.  Following the
terminology of \citet{car08}, the blue circles reproduce the
Primordial population, while the cyan ones give the Intermediate and
Extreme ([O/Fe]$<-0.4$) populations. We emphasize that without the
modification in the chemical composition of the FG stars discussed
above (cf. sect. \ref{subsec:ona}) the Extreme population and part of
the Intermediate ones could not be reproduced. In any case, stars with
[Na/Fe]$>0.5$ remain unexplained, and suggest that a further parameter
is at work. In order to obtain a better fit of the data points, we
could have increased the sodium produced in the mass range 6.5 --
9 $\msun$ to [Na/Fe]$\sim$0.7. This procedure has two negative
implications: first, it can not explain, in any case, the six points
at [Na/Fe]$>$0.7 having upper limits for the oxygen values. In
addition, such an extrapolation, implying an increase of the sodium
yield with increasing initial mass, is not compatible with our present
understanding of the super--AGB evolution (see
Sect. \ref{subsec:supagb}). We prefer then to postpone a better
solution of this problem to when more data (e.g., when the oxygen
abundance for these stars will be measured) and better models will be
available.  The imperfect reproduction of these (few) data is not an
argument to believe that the origin of the gas forming these SG stars
must be looked for somewhere else than among super--AGBs. In fact, no
other source can be able to deal, at the same time, with the
reproduction of the Mg--poor, Al--rich stars \citep{decressin2007},
and with the extreme but homogeneous helium content that is necessary
to reproduce the triple main sequence \citep[][and
Fig. \ref{fig:2808}, top right panel]{der08,dc2008,ren08}.  In
summary, we regard the fit of NGC 2808 data as very satisfactory, in
spite of this remaining problem.

The three [Al/Fe]-rich stars in the upper-left panel of Fig.
\ref{fig:2808} coincide with the three magenta arrows in the
[O/Fe]-[Na/Fe] diagram. In both diagrams these UVES data 
\citep{carretta2009b} show a very
``clumpy'' anticorrelation, with only Primordial and Extreme populations
present, and a gap in the middle where no star is found. On the other
hand, our model, giving rise to stars occurring throughout the whole
O-Fe anticorrelation span, necessarily produces a ``continuous'' distribution
of stars in the [Mg/Fe]-[Al/Fe] diagram. We believe that the gap in the
Mg-Al anticorrelation displayed by the UVES data is not real (as indeed
shown by the FLAMES/GIRAFFE data \citep{carretta2009a} 
in the [O/Fe]-[Na/Fe] plane), and we predict
that an Intermediate population in the [Mg/Fe]-[Al/Fe] diagram will be observed
in the next future.

Finally, the top-right panel of Fig. \ref{fig:2808} shows the helium
stellar distribution. Three stellar populations are clearly
present. The most He-rich stellar group ($Y>0.35$) originates from the
FG ejecta before the pristine gas enters the GC, while the
Intermediate population ($0.3\la Y<0.35$) forms after the accretion of
the original gas.  The ratio between Extreme and Intermediate He-rich
stars is 0.56, as observed. Contrary to the abundance of the FG stars,
which is simply imposed, the value of the ratio between Extreme and Intermediate
stars is a genuine result of our model.

\subsection{M4 (NGC 6121)}
\label{sec:m4}
The globular cluster M4 has a mass $M=6.3\times 10^4$ $\msun$
\citep{man91} and its color-magnitude diagram does not show any of the
fingerprints of the presence of multiple populations found in the most
massive clusters. However, the O-Na anticorrelation is present in this
cluster as in all the Galactic GCs. Instead, contrary to NGC 2808,
there is no Mg-Al anticorrelation in M4 \citep{mar08}. As the stellar
evolution must be the same in all the GCs, our aim is to fit M4 data
adopting the same chemical composition of the AGB ejecta utilized in
modelling NGC 2808. However, we must assume a different composition of
the pristine gas. This can be understood by looking at the
observational data in Fig. \ref{fig:2808} and \ref{fig:m4}. The FG
oxygen, sodium, magnesium and aluminium for NGC~2808 are [O/Fe]$\sim
+0.4$, [Na/Fe]$\sim 0.0$, [Mg/Fe]$\sim +0.28$\ and [Al/Fe]$\sim
+0.15$.  In M4 we have [O/Fe]$\sim +0.5$, [Na/Fe]$\sim +0.1$,
[Mg/Fe]$\sim +0.5$ and [Al/Fe]$\sim +0.4$.  Therefore, with respect to
the input values for NGC 2808, in the present model we adopt O and Na
abundances 25\% higher, and Mg and Al abundances 67\% and 78\% higher,
respectively. As a consequence, we should also consider that the {\it
  initial} abundance of \neo20, another $\alpha$ element, may have
been larger in the pristine gas. Following the test computations
described in section 2.2, also the sodium yield of the AGBs will be
larger, and we can reasonably increase by a factor 0.2 dex the ratio
[Na/Fe] of the ejecta of the AGB of any mass.  Notice that this last
change in the yield table is only applied to the M4 case, due to the
different initial chemistry of its FG.

\begin{figure}    
\centering{
\includegraphics[width=8cm]{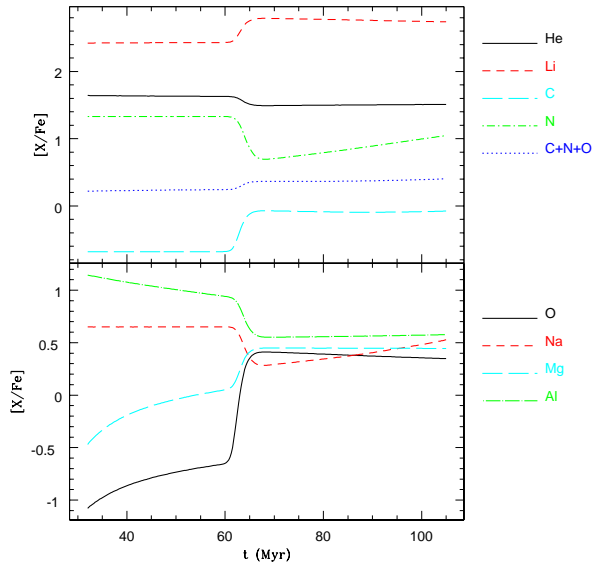}
}
\caption{Time evolution of the chemical abundances in the ISM of 
the M4 model.
}
\label{fig:chmm4} 
\end{figure}

For this model of M4 we set the parameters $(t_{\rm ac,7},t_{\rm
  end,7},\tau_7,\rho_{\rm pr},\rho_{\rm
  *,FG},\nu,x)=(6.5,10.5,0.91,9.4,0.1,0.7)$. The results are
illustrated in Fig. \ref{fig:chmm4} and Fig. \ref{fig:m4}. The former
figure shows the evolution of the ISM chemical abundance of
several elements; in particular, it can be seen that the variation of
C+N+O abundance is not larger than 0.1 dex, as required by the
observations \citep{yong2009}. Figure \ref{fig:m4} can be compared to
the analogous Fig. \ref{fig:2808} relative to NGC 2808 to understand
the different role played by various parameters in the two cases.

\begin{figure}    
\centering{
\includegraphics[width=8cm]{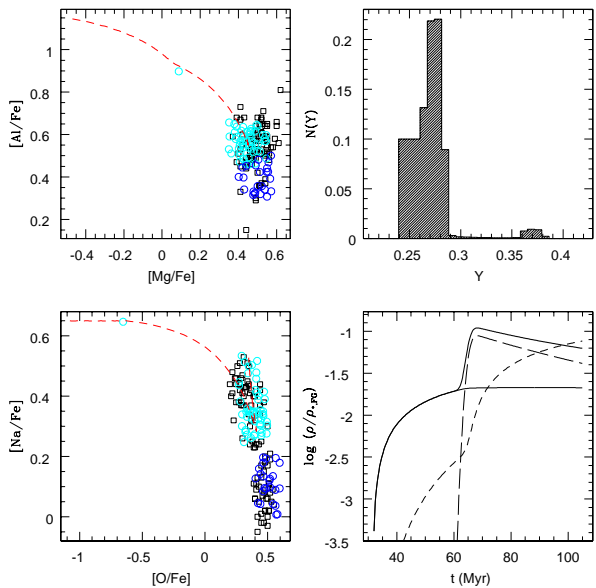}
}
\caption{As Fig. \ref{fig:2808}, but for M4. Data are from
  \citet{mar08}. In this case, for a better comparison with the data,
  the error on [Na/Fe] is taken 0.05 dex instead of 0.1 dex as in the
  case of NGC 2808.  }
\label{fig:m4} 
\end{figure}

Our model shows that the relative amount of pristine gas must be
larger than in NGC 2808.  The reason is illustrated in the bottom-left
panel of Fig. \ref{fig:m4}. The Na-O anticorrelation is less extended
than in NGC 2808; in particular, all the stars in M4 exhibit a
positive ratio [O/Fe], while in NGC 2808 stars are found with ratios
as low as [O/Fe]$\sim -1$. In order to reduce the fraction of this
kind of stars we set $\nu=0.1$ so to minimize the SFR in the
beginning, before the accretion of pristine gas, when the ISM is
composed only of O-poor AGB ejecta. After the pristine gas accretion
and the consequent increase of the ISM density, the SFR becomes
significant despite the low value of $\nu$. Thus, a large fraction of
SG stars form from an ISM strongly ``diluted'' by the pristine O-rich
gas.

In order to form the observed Na-rich stars, the model for M4 also
requires a value of $t_{\rm end}$ larger than that of NGC 2808. By
extending the SF duration, Na-rich stars can form at late time when
the pristine gas has been strongly depleted by SF and the ISM is
mainly composed of the Na-rich stellar ejecta of stars with $M<6$
$M_{\odot}$.

The chemical evolution of the GC is summarized by the dashed line in
the bottom-left panel of Fig. \ref{fig:m4}. The model starts at the
O-poor extreme and moves rightward relatively fast and in a low SFR
regime, so that only $\sim 3$\% of the total SG population form during this
excursion. The trajectory has a cusp at [O/Fe]$\sim 0.4$ where
then develops a rather vertical branch due to the reduction of
pristine gas and the increment of Na-rich ejecta in the ISM
``mixture''. The longer length of time spent along this branch and,
more importantly, the larger SFR during this stage determine the
clustering of the SG stellar population in this area of the Na-O
plane.

The amount of pristine gas accreted by the GC is a key parameter to
fit the observed data {\it both} in the [O/Fe]-[Na/Fe] and the
[Mg/Fe]-[Al/Fe] diagrams. An increment of pristine gas causes the
abundances of SG stars to approach those of FG stars and, in
particular, reduces the differences in the [Na/Fe] and [Al/Fe] ratios
between Primordial and Intermediate stars. The Na abundance is,
however, more sensitive to this effect: a model of ours (not shown
here) reveals that, after trebling the accreted gas, the Al abundances
are still consistent with the data, while stars with [Na/Fe]$>0.4$ can
not be reproduced.

Finally, from the top-right panel of Fig. \ref{fig:m4} we note that
the helium abundance of the SG Intermediate population is not much
larger than that of the FG one, and the two populations, although
still distinguishable, blend together in the HDF. This small helium
spread can be revealed in careful observations of the main sequence
width \citep[see, e.g., the analysis by][for the cluster
NGC~6752]{milone2010} and for its possible consequences on the
horizontal branch (HB) morphologies \citep{dc2008}. The HB of M4 shows
signs of bimodality \citep{norris1981} but a careful representation of
the data is very tricky, as it requires use of models including also
the CNO abundance enhancements of the SG \citep[see the discussion
by][concerning the bimodal distribution in the HB of
NGC~1851]{salaris2008}. \cite{dantona2009} notice that the blue side
of the HB of M4 might be slightly overluminous with respect to the
blue HB of NGC~1851, and tentatively attribute this feature to
possible helium enhancement in these stars. This would be consistent
with our helium distribution in M4.

The Extreme population is still present, but is only 2\% of the whole
GC stellar population. As observationally this population is lacking
in the clusters that are not very massive, this may be an indication
that the formation of the Extreme population is inhibited below some
critical mass. Alternatively, given its paucity, we can guess that
this population exists but is not yet observed\footnote{Notice however
  that a very helium rich, even small, population would be
  photometrically very evident, as it would populate the extreme blue
  horizontal branch\citep{dc2008}.}.

In any case, as shown in sect. \ref{subsec:space}, it is enough to
change somewhat one or more parameters to create or remove small
structures in the helium distribution function. Thus, at this stage,
no firm conclusion can be drawn on the Extreme population we found, or
on the detailed helium spread of the Intermediate population. More
realistic models are planned, in which the chemical machinery
developed in this paper will be implemented in a hydro-code.

\section{Conclusions}
\label{sec:conc}
In this paper we have studied the origin of the chemical patterns
which have been observed in many globular clusters and which are
considered the spectroscopic fingerprints of the presence of multiple
stellar populations.  Specifically, in our investigation we have
focussed our attention on the O-Na and Mg-Al anticorrelations and the
helium distribution function.

In our model we have assumed that AGB stars are the polluting source
of gas with the anomalous abundances of light elements observed in
globular cluster second-generation stars.  Our chemical framework is
based on a one-zone model following the formation of SG stars from a
mix of ejecta of AGB stars and pristine gas.

We have carried out a large number of simulations to explore the
dependence of the SG chemical properties on the parameters
characterizing the star formation process and the dynamics of the
involved pristine gas. Finally, we have used our
framework to model the observed chemical patterns in the Galactic
clusters NGC 2808 and M4.

The main results of our study are the following:

\begin{enumerate}
\item The current stellar models provide AGB ejecta in which both Na
  and O decrease with increasing stellar mass. In order to reproduce
  the observed O-Na anticorrelation, the gas from which SG stars form
  must be diluted with O-rich, Na-poor pristine gas (see
  e.g. Figs. \ref{fig:refmod} and \ref{fig:refchm}).
\item The amount of pristine gas involved in the SG formation process,
  the timescales driving the dynamics of such gas, and the star
  formation efficiency play a key role in determining the extension of
  the O-Na anticorrelation and the fraction of extreme Na-rich/O-poor
  stars.
\item The helium abundance distribution function is correlated with
  the distribution of stars in the O-Na plane. Extreme Na-rich/O-poor
  stars are also those with an extreme He enrichement. Our models
  predict that all the clusters with a very extended O-Na
  anticorrelation should also host a population of He-rich stars.
\item Our models show that the extension of the O-Na anticorrelation
  is closely correlated with that in the Mg-Al plane; Na-rich/O-poor
  stars have also high Al and low Mg abundances.
\item We have used our framework to build specific models for two
  prototypical Galactic globular clusters: NGC 2808, a massive cluster
  that hosts a SG population characterized by a very extended O-Na
  anticorrelation and includes a very He-rich population, and M4, a
  low-mass cluster with a significantly less extended O-Na
  anticorrelation that does not include extremely Na-rich O-poor stars
  and for which there is no photometric evidence of a He-rich
  population. Despite the significant differences in their chemical
  patterns, in both clusters a significant fraction of stars belong to
  the SG \citep[50 per cent in NGC 2808,] [and 65 per cent
  in M4 \cite{carretta2009a,mar08})]{carretta2009a}.


  Our models successfully reproduce the differences in the O-Na
  anticorrelation observed in these two clusters, the distribution of
  stars in the Mg-Al plane in M4 and predict an extended Mg-Al
  anticorrelation for NGC 2808. Al and Mg abundances for NGC 2808 have
  been determined with UVES observations only for a small number of
  either Na-rich/O-poor or Na-poor/O-rich stars. As predicted by our
  model, these stars populate only the Al-rich/Mg-poor and the
  Al-poor/Mg-rich regions of the Mg-Al plane. According to our model,
  future determinations of Al and Mg abundances of targeting
  stars with intermediate Na and O should lead to intermediate Al and Mg
  abundances filling the extended Al-Mg anticorrelation.

\item The helium distribution, including the extreme population formed
  directly from super--AGB ejecta  is well reproduced by our model for
  NGC~2808 (see Fig. \ref{fig:2808}). In clusters like M4, in which a
  larger dilution with pristine matter is necessary to model the O-Na
  and Mg-Al patterns, a large helium dispersion is not required, but a
  small helium spread should generally be present (see
  Fig. \ref{fig:m4}).
\end{enumerate}

Our investigation has shed light on the key chemical and
hydrodynamical ingredients determining the formation of the chemical
patterns observed in globular clusters.
The results presented here are to be the starting point informing
further study based on full hydrodynamical simulations. Several issues
will require further investigation. Specifically, as for the stellar
evolution models, we have shown that our conservative choice of an
educated linear and monotonic extrapolation for the sodium and oxygen
abundances in the mass range from 6.5 to 9 $\msun$ reproduces the
general trend of the O--Na anticorrelations of different clusters,
when also the differences in neon abundances in the FG of different
clusters are taken into account. In NGC~2808, however, we can not
reproduce the (few) very large sodium values of some stars for which
only upper limits on the oxygen abundance are available (see
Fig. \ref{fig:2808}). Another parameter is probably at work here, and
only further investigation into the super--AGB phase will shed light on this
issue.

The lack of any evidence of a significant metal enrichment in most
clusters hosting multiple stellar populations implies that neither
ejecta from FG nor SG supernovae  is involved in the chemical
enrichment of matter from which SG stars form. While in our model SG
formation starts after the end of the FG SN II epoch, it is to be
further clarified whether the lack of metal enrichment from SG
supernovae is due to SG forming with IMF truncated at $M<9$ $\msun$
or whether this is a consequence of a more complex gas dynamics to be
further explored with full hydrodynamical simulations.

Finally, additional full hydrodynamical simulations will be needed to
clarify the source of pristine gas and the accretion mechanism.
Different processes have been considered in the literature, such as
accretion from a diffuse surrounding medium
\citep[e.g.][]{limu07,pflkro09} or interaction with molecular clouds
\citep{bekmac09}. Conditions for an effective gas collection within
the cluster turn out to be rather specific. Instead, our model
requires a mechanism general enough to work in all the clusters and
able to take into account the cluster-to-cluster differences in the
dynamics and the amount of pristine gas involved in the SG formation
process.

\section*{Acknowledgments}
We are grateful to E. Carretta and A. Renzini for useful disussions.
A.D. acknowledges financial support from italian MIUR through grant PRIN
2007, prot. 2007JJC53X. F.D. and P.V. have been supported through PRIN
MIUR 2007 ``Multiple stellar populations in globular clusters: census,
characterization and origin" (prot. n. 20075TP5K9).
E.V. and S.M. were supported in part by NASA grant NNX10AD86G.

\bibliographystyle{mn2e} 
\bibliography{gc_ref}
\label{lastpage}
\end{document}